\documentclass[%
aps,pra,
superscriptaddress,
 mph,%
 amsmath,amssymb,
 reprint,%
]{revtex4-2}

\usepackage{graphicx}
\usepackage{dcolumn}
\usepackage{bm}
\usepackage{color}
\usepackage{epsfig}
\usepackage{cleveref}
\usepackage{bm}
\usepackage{comment}
\usepackage[normalem]{ulem}

\usepackage{amsfonts}
\usepackage{float}
\usepackage{multirow}
\usepackage{pgfplots}

\usepackage{fancyhdr}

\begin{document}

\preprint{}

\title{Propagating single photons from an open cavity: Description from universal quantization}
\author{A. Saharyan}
\author{B. Rousseaux}
\affiliation{Laboratoire Interdisciplinaire Carnot de Bourgogne, CNRS UMR 6303, Universit\'e de Bourgogne,
BP 47870, 21078 Dijon, France}
\author{Z. Kis}
\affiliation{Wigner
  Research Center for Physics, Konkoly-Thege Mikl\'os \'ut 29-33, 1121
  Budapest, Hungary}
\author{S. Stryzhenko}
\affiliation{Institut für Angewandte Physik, Technische Universität Darmstadt, \\
Hochschulstraße 6, 64289 Darmstadt, Germany}
\author{S. Gu\'erin}
\email{sguerin@u-bourgogne.fr}
\affiliation{Laboratoire Interdisciplinaire Carnot de Bourgogne, CNRS UMR 6303, Universit\'e de Bourgogne,
BP 47870, 21078 Dijon, France}

\date{\today}

\begin{abstract}
Over the last decades, quantum optics has evolved from high quality factor cavities in the early experiments toward new cavity designs involving leaky modes. Despite very reliable models, in the concepts of cavity quantum electrodynamics, photon leakage is most of the time treated phenomenologically. Here, we take a different approach, and starting from first principles, we define an inside-outside representation which is derived from the original true-mode representation, in which one can determine effective Hamiltonian and Poynting vector. Contrary to the phenomenological model, they allow a full description of a leaking single photon produced in the cavity and propagating in free space. This is applied for a laser-driven atom-cavity system. In addition, we propose an atom-cavity non-resonant scheme for single photon generation, and we rigorously analyze the outgoing single photon in time and frequency domains for different coupling regimes. Finally, we introduce a particular coupling regime ensuring adiabatic elimination for which the  pulse shape of the outgoing single photon
is tailored using a specifically designed driving field envelope.

\end{abstract}

\maketitle

\section{Introduction}
Single photons  are nowadays key elements in  quantum technologies, as
quantum  networking  for  distributed computation,  communication  and
metrology {\cite{nielsen00,Bouwmeester,Cacciapuoti1,Cacciapuoti2,Qinternet,Kimble2008,ciracQuantumOpticsWhat2017a}.}  Sources producing single  photons have
been  widely developed  \cite{Lounis_2005,santori}. {Their quantization and treatment as wave  functions} in  connection with  a corpuscular
viewpoint       have      been       debated       until      recently
\cite{Bialynicki,Sipe,Smith}. From a practical  point of view, one can
for  instance mention {the need for such a description} in quantum  cryptography \cite{Gisin}
over the  use of  attenuated laser pulses  for making the  security of quantum key  distribution device-independent, or  for extending quantum
communication over very long distances \cite{Sangouard,barrettMultimodeInterferometryEntangling2019a,Guerin_2021,Cacciapuoti1,Cacciapuoti2}. An envisioned
quantum network makes use of single photon wavepackets as carriers of quantum information  (encoded for instance in the polarization state
giving flying qubits) to map  the states between distant quantum nodes
{\cite{Cacciapuoti1,Qinternet,Kimble2008,Fan2009}}, such as individual   atoms   in  cavity quantum electrodynamics (cavity QED)  \cite{cQED1,cQED2,McKeever,Wilk,Ritter2012,Martin,Boozer,cQED4, Dilley,multilayer,Vahala2003,Dawson,PhysRevA.97.043827},            atomic            ensembles
\cite{Gorshkov,Choi2008},  trapped  ions  \cite{Keller2004,Piro2011}, or  spins  in
quantum dots \cite{Yao,Ding}. One key  point is to control the node-photon
interfacing, {i.e. to have control over the produced photon frequency, bandwidth, and temporal shape such that} {the node can send},  receive, store  and
release photonic quantum information { \cite{Fischer2018scatteringintoone,PhysRevX.4.041010,PhysRevA.101.013808,PhysRevA.96.043822,PhysRevApplied.8.054015,Keller2004}. This control is in general achieved by
control laser pulses}. Recent  studies have investigated the control of
the  shape of  the single photon wavepackets in  $\Lambda$-atoms  by a
resonant  stimulated  Raman process {\cite{Vasilev_2010,Nisbet_Jones_2011,Alvarez_2022, Rephaeli_2012}}  in order  for
instance  to  improve  the   impedance  matching  of  the  atom-photon
interface  \cite{cQED4,Dilley}.  The  possible  production  of  more  complex
traveling     photonic      states     featuring     $N>1$     photons
\cite{Brown,Eleuch,Amniat-Talab,Gogyan}  can be  envisioned for  the  transport of
complex information. For instance,  the delays and relative amplitudes
between the  pulse-shaped individual photons offer a  large variety of
encoding,  which generalizes  the  possibility of  producing a train  of
well-separated pulses \cite{Gogyan_A}.

Cavity QED, the theory of atoms coupled to single mode cavities, is nowadays well-known \cite{gardiner00,Loudon,Breuer,Qinternet}. More recently, transposition of cavity QED to leaky cavities has, however, led to misinterpretations in nanophotonics \cite{Tserkezis_2020}. These issues are mainly due to a misuse of the models derived for high quality factor (high-Q) cavity QED experiments, as opposed to full quantized treatment \cite{DorierGuerinJauslin,PhysRevB.93.045422}. Indeed, they were derived with a phenomenological system-reservoir approach to describe the cavity leakage, where a flat continuum {with a constant coupling} is assumed for the reservoir, perturbatively broadening the cavity resonances {(see e.g. comments in \cite{Fischer2018scatteringintoone})}. Derivation from first principles is highly desirable, even for the case of high-Q cavities, since it will provide the community with clearer aspects in the limits of the applicability of these well-known models.

{In this paper, we establish such a derivation: starting from the universal quantization of the true modes in a semi-infinite system composed of a perfectly reflecting boundary and a semi-transparent mirror, we determine effective models for a laser-driven atom in a cavity and characterize the resulting propagating photon field.} To this aim we define and connect concepts, namely photon fluxes, input-output operators, quantized Poynting vector, effective master equation, photonic wavepackets and states {to this concrete physical situation \cite{gardiner00,Loudon,Breuer}}. We derive an inside-outside representation from first principles, allowing to characterise the leaking photon in time as well as in frequency domain. {We analyze the validity of different representations, i.e., the true mode picture, the inside-outside representation, and the pseudo-mode picture, by comparing the dynamics obtained in each representation}. We apply the model for a non-resonant scheme in a three-level atom trapped in a cavity and show that it allows a direct and simple way to design the photonic wavepacket on demand. This is obtained for a particular coupling regime, which ensures single photon production without populating the cavity state. This leads to the production of a single photon with broad bandwidth, which can be of advantage when coupling {photon states} with materials of distinct resonances.

This  paper is organized as follows: In  Section~\ref{sec:model},  we  introduce {different representations: true mode, inside-outside, and pseudo-mode picture.} From the equivalence of {true mode and inside-outside representations,} we write the cavity-reservoir coupling function \cite{dutra2005cavity}, which is later used to analyze the dynamics of the leaking photon. The explicit form of this coupling function allows one to derive the standard input-output formulation {\emph{without applying any a priori approximations} \cite{Breuer,Scala,Scala_2007,Giorgi_md} leading to false mathematical justifications of the Markov approximation} (see e.g. comments in~\textcolor{black}{Appendix \ref{app:delta_ev}}).
We connect the
photon flux, corresponding to the propagation of the photonic state in
free  space leaking  from  the cavity  to  the quantum  average of  a
reservoir  photon number operator,  in the  Heisenberg picture,
using the  quantized Poynting vector {derived from the true mode representation}. The  condition of correspondence
of this reservoir photon number operator to the standard output photon
number operator is derived. We  next establish that the photon flux is
proportional  to  the quantum  average  of  the  cavity photon  number
operator   in  the   condition   of  an   initial  ground   state
reservoir. {The master  equation, which  allows one to  determine the state of the atom-cavity system, is finally derived. In section
\ref{sec:prod_sing},  we apply  the  derived  model  to the production of shaped single photon wavepackets, using a non-resonant laser pulse scheme for a three-level atom in a ``$\Lambda$'' configuration inside of a high-Q cavity. We provide a summary in Section \ref{sec:conclusion}.}

\section{\label{sec:model} Derivation of the model}

In  this  Section,  we  introduce {different representations} for deriving the dynamics of an atom trapped in an optical cavity and driven by a classical field. {In particular, we introduce the inside-outside representation, which} assumes separation between the modes of the inside and the outside of the cavity. We {analyze the validity of this separation} by comparing the dynamics obtained via inside-outside representation to the one obtained from the universal quantization of the true, unseparated modes of Maxwell's equations in a one-dimensional semi-infinite space \cite{dutra2005cavity,vogel2006quantum, multilayer}.  We then connect  the  photon  flux  \cite{Blow_1990,Law},
corresponding to the  propagation of the photonic state  in free space
leaking from the cavity, to  the quantum average of a reservoir photon
number operator, in the Heisenberg picture, constructed with an
integrated   reservoir   operator. {We derive the quantized Poynting vector from the true mode representation, which we then write in terms of the reservoir operators corresponding to the inside-outside representation.}  We derive  the condition of correspondence  of this reservoir  photon number  operator to  the standard output photon number operator derived in the input-output formulation
\cite{gardiner00}. We next  establish that the photon flux  is proportional to
the  quantum average  of the  cavity photon  number operator  when the
reservoir  is initially  in  the ground  state \cite{Gogyan_A}.   We
finally derive the master  equation \cite{Breuer,gardiner00,gheri1998photon} by tracing out the reservoir degrees of freedom, which allows one  to determine the state of the {atom-cavity system, necessary to obtain the quantum averages describing the relevant physical observables.}

\subsection{Hamiltonian in the Schr\"odinger picture}

We consider a single $\Lambda$-atom with ground $|g\rangle$, metastable $|f\rangle$ and excited $|e\rangle$ states
trapped  in  a cavity, {which is designed to sustain a field of wavelength $\lambda_{c}$ and frequency $\omega_{c}$}. {The $|f\rangle\leftrightarrow  |e\rangle$ transition, with frequency $\omega_{ef}$ and dipole moment ${d}_{fe}$ is assumed to be nearly resonant with a cavity mode of area ${\cal A}$ and length $L$, {with the detuning $\Delta_{c}=\omega_{ef}-\omega_{c}$}; the $|g\rangle\leftrightarrow  |e\rangle$ transition, with frequency $\omega_{eg}$ and dipole moment $d_{ge}$ is assumed to be independently driven by a classical laser field ${\cal E}(t)\cos(\omega_{0}t+\varphi)$, corresponding to the time-dependent Rabi frequency $\Omega(t) = -{\cal E}(t)d_{ge}/2\hbar$, {with a detuning $\Delta = \omega_{eg}-\omega_{0}$}. In this work, we assume the mirrors of the cavity to be large enough, so that the spontaneous emission of the atom in modes propagating in directions {perpendicular to the optical axis} are neglected. Thus, we consider the cavity to be a one-dimensional Fabry-P\'erot resonator.}

\subsubsection{\label{sec:universal} {True mode representation}}

\begin{figure}[!ht]
\begin{center}
\includegraphics[width=.45\textwidth]{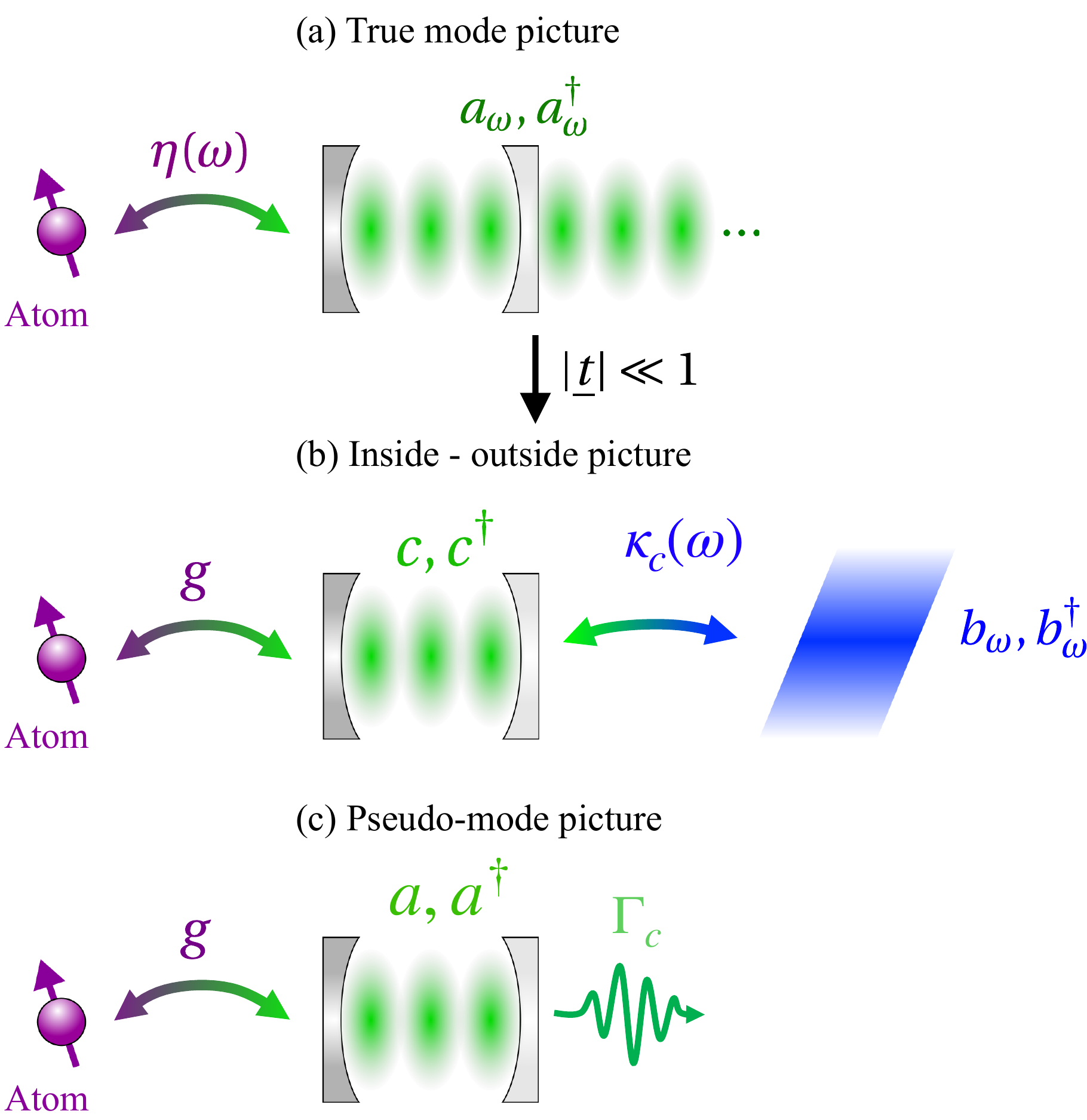}
\caption{(a) True mode picture corresponding to the modes obtained from universal quantization, where the cavity is treated as part of the environment. In such a representation, the atom is coupled to the universal modes $a_{\omega}$ with a frequency-dependent coupling strength $\eta(\omega)$. (b) For cavities with sufficiently small transmission it is possible to {red}{approximately separate modes into} cavity modes (inside) and reservoir (outside), with an effective frequency-dependent coupling $\kappa_{c} (\omega)$. In this picture, the atom couples {mainly} to the cavity mode $c$,  {to which it is resonant (or near resonant)}, with coupling strength $g$. (c) Unlike the two other representations, in the pseudo-mode picture, the reservoir is eliminated and accounted for via the cavity decay rate $\Gamma_{c}$, in a non-Hermitian description.}
\label{fig:mode_representations}
\end{center}
\end{figure}

The universal quantization procedure \cite{dutra2005cavity,vogel2006quantum, multilayer, Knoll, murray1978laser}  {is a derivation from first principles, which} allows the treatment of cavity as part of the environment and the derivation of true (exact) modes for such a closed system (see Fig.~\ref{fig:mode_representations}). Here we consider the same physical situation as the one in \cite{dutra2005cavity}, where the cavity is delimited by a perfect mirror on the left and a semi-transparent mirror on the right, the latter being made of a single layer of a dielectric material. The length of the dielectric layer is considered to be negligible {with respect to the cavity length}. By introducing an atom in such a cavity, we can write the Hamiltonian for the full system ${\mathcal{A}\oplus \cal E}$ in a rotating  frame defined by the unitary operator ${U}_{\rm
    RW}= \exp \left(\textcolor{black}{-}i\omega_0t \right) \sigma_g + \sigma_e + \sigma_f $:
    
\begin{subequations}
\label{Hamiltonian_universal}
\begin{align}
&\tilde{H}(t)={H}_{A}(t)+{H}_{\text{int}}+{H}_{E}\\
\label{eq:atom_hamiltonian}
&{H}_{A}(t)={\hbar \left(\Delta-\Delta_{c}-\omega_{c}\right)}\sigma_{f}+\hbar \Delta \sigma_{e}+\hbar\Omega \left(\sigma_{ge} +\sigma_{eg}\right),\\
&{H}_{E} = \int_0^{+\infty} d\omega \; \hbar \omega a^{\dagger}_{\omega}a_{\omega},\\
&{H}_{\text{int}} = i\hbar \int_0^{+\infty}  d\omega \; \left( \eta(\omega) a_{\omega} \sigma^{\dagger } - \eta^{\ast }(\omega) a^{\dagger}_{\omega} \sigma \right),
\end{align}
\end{subequations}
where ${H}_{A}\equiv{H}_A(t)$  denotes  the  atomic Hamiltonian {in rotating wave approximation (RWA)}. {Here, we have     introduced the  atomic
operators  $\sigma_{k\ell}\equiv  |k\rangle\langle  \ell|$, $\sigma_{k}\equiv\sigma_{kk}$
and  $\sigma\equiv\sigma_{fe}$.} ${H}_{E}$ describes the environment with the cavity as part of it. {Operators ${a}_{\omega},  {a}^{\dagger}_{\omega}$ are the annihilation and creation operators for the true modes, satisfying the commutation relation:
\begin{eqnarray}
[{a}_{\omega},{a}^{\dagger}_{\omega'}]&=&\delta(\omega-\omega').
\end{eqnarray}
${H}_{\text{int}}$} represents the interaction between the atom and the {structured} environment, with the coupling factor
\begin{subequations}
\label{eq:coupling_full}
\begin{align}
\label{eq:eta_real}
\eta(\omega) ={}& { i\sqrt{\frac{\omega}{\hbar \epsilon_{0}\pi c \mathcal{A}}}d_{fe}e^{i\frac{\omega}{c}L}\sin{\left(\frac{\omega}{c}(x_{A}+L)\right)}T(\omega)}\\ \nonumber
 \approx{}& i\sqrt{\frac{\omega}{\hbar \epsilon_{0} L \mathcal{A}}}d_{fe}e^{i\frac{\omega}{c}L}\sin{\Big( \frac{\omega}{c}\left( x_{A}+L\right)\Big)}\\
&\times \sqrt{\frac{\Gamma_{c}}{2\pi}}\frac{1}{\omega-\omega_{c}+i\frac{\Gamma_{c}}{2}},
\label{eq:coupling_full_Lor}
\end{align}
\end{subequations}
{where $T(\omega)$ is the single layer cavity response function and $x_{A}$ is the position of the atom. For a cavity with sufficiently high reflectivity} the response function can be represented as a sum of mode-selective Lorentzian functions, whose width $\Gamma_{c}$ (decay rate of the cavity) is much smaller than the spacing $\Delta_{\omega}=\pi c/L$ between the neighbouring resonances {(high finesse cavity)} \cite{dutra2005cavity,vogel2006quantum,multilayer}  {(see the details in Appendix~\ref{app:Lorentzian_separation}).} For a single layer {partially transparent} cavity {mirror}, high reflectivity can be achieved by assuming a fictitiously large refractive index of the dielectric material, while for more realistic models high reflectivity implies a cavity  {mirror} made of a large number of dielectric layers \cite{multilayer}. {In the literature, these assumptions usually correspond to the high-${Q}$ cavity limit. However, as we show below, the high-$Q$ assumption by itself is not sufficient, by definition. Instead, we should require high finesse cavity, which satisfies all the conditions necessary to perform the above approximations.}
 
{In Eq.~\eqref{eq:coupling_full_Lor}, we assume that the atom couples to a single mode ($\omega_{c}$), reducing the sum of Lorentzians into a single one}. The Hamiltonian \eqref{Hamiltonian_universal} with the coupling \eqref{eq:coupling_full} describe a true-mode representation, i.e., with continuous frequencies, but with a structured reservoir \cite{pseudomodes}. {In the following we break these true modes into inside and outside modes, which describe the cavity and the reservoir separately.}

\subsubsection{\label{sec:in_out} Mode separation into inside-outside modes}

\begin{figure}[!ht]
\begin{center}
\includegraphics[scale=0.45]{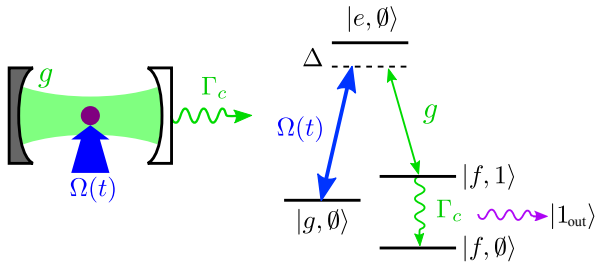}
\caption{ Atom-field  interaction  in  the  cavity:  (left
  panel)  a  single  $\Lambda$-atom   is  driven  by  an  external
    classical  laser field  of  Rabi frequency  $\Omega$, and  a
  quantized  cavity   field  with  coupling   strength  $g$.
  (Right  panel) The  fields  are in  two-photon  resonance {($\Delta=\Delta_{c}$)},  the
    one-photon detuning  is $\Delta$.   Initially the atom  is in
    the ground state $|g\rangle$. In the course of the excitation process, one
  photon is taken from the  laser field and transferred to the cavity,
  which   eventually  leaks   out   of  the   cavity  through   a
  semi-transparent  mirror   characterized  by  the   decay  rate
  $\Gamma_c$.}
\label{1atc}
\end{center}
\end{figure}

We now consider an approximately equivalent model to the one obtained previously via splitting the modes $a_{\omega}$ into two parts: cavity modes $c$ ({inside}) and the continuum of reservoir modes $b_{\omega}$ ({outside}) (Fig.~\ref{fig:mode_representations}~{(b)}). The derivation is formally shown in \cite{dutra2005cavity}, exhibiting an error of order $\mathcal{O}(|{\underline{t}}|^{2})$, where ${\underline{t}}$ is the transmission rate of the (single layer) mirror. {This representation} can be interpreted as replacing the semitransparent mirror by a perfect one, forming a perfect cavity  ($\mathcal{C}$), which is coupled to the reservoir ($\mathcal{R}$) \cite{dutra2005cavity} (see Fig.   \ref{1atc} for  the
  coupling scheme of the atom with the cavity). We refer to it as the inside-outside representation. The {RWA}
Hamiltonian           of           the           full           system
${\mathcal{A}\oplus\mathcal{C}\oplus\mathcal{R}}$
reads, in  the Schr\"odinger picture:
\begin{subequations}
\label{Hamiltonian}
\begin{eqnarray}
&&{H}(t)= {H}_{A}(t)+{H}_{AC}+{H}_{C}+{H}_{RC}+{H}_{R}\\
&&{H}_C=\hbar\omega_c c^{\dagger}c,\\
&&{H}_{AC}=\hbar g \big(c^{\dagger}\sigma +\sigma^{\dagger }c\big),\\
&&{H}_R=\int_0^{+\infty}\hspace{-1em} d\omega\,\hbar\omega\,b_{\omega}^{\dagger}b_{\omega},\\
&&{H}_{RC}=i\hbar\int_0^{+\infty}\hspace{-1em} d\omega\,\big( \kappa_{c}(\omega) b_{\omega}^{\dagger}c
-\kappa_{c}^{\ast}(\omega) c^{\dagger}  b_{\omega}\big),
\end{eqnarray}
\end{subequations}
with the atom-cavity coupling factor
$g=-d_{fe}\sqrt{\omega_c/\hbar\epsilon_0 L \mathcal{A}}$      (one-photon     Rabi
frequency), assuming the atom is localized at the field maximum. The coupling factor $\eta(\omega)$ for the {true mode picture} (Eq.~\eqref{eq:coupling_full_Lor}) can then be approximated as \cite{multilayer}
\begin{eqnarray}
\label{eq:eta_approx}
\eta(\omega)\approx {\hat{\eta}(\omega)=}-ig\sqrt{\frac{\Gamma_{c}}{2\pi}}\frac{1}{\omega-\omega_{c}+i\frac{\Gamma_{c}}{2}}.
\end{eqnarray}
{This form allows the direct derivation of the pseudo-mode representation described below [see Eq.~\eqref{eq:effective_Hamiltonian}]}.

In   Eq.~\eqref{Hamiltonian},
$H_{A}(t)$ is the same as Eq.~\eqref{eq:atom_hamiltonian},  ${H}_C$   is  the   free   cavity
Hamiltonian, ${H}_{AC}$  describes the coupling  between the atom
and  the cavity, ${H}_R$  is the  free reservoir  Hamiltonian, and
${H}_{RC}$    describes   the    coupling between     the    {empty cavity and the free reservoir.}  
The reservoir  annihilation and creation operators
$b_\omega,b_\omega^\dagger$   satisfy   the   commutation
relation:
\begin{equation}
[b_{\omega},b_{\omega'}^{\dagger}]=\delta(\omega -\omega').
\end{equation}
 
The cavity-reservoir coupling function $\kappa_{c} (\omega)$ can be evaluated in the limit of small transmission and {near resonance} as \cite{dutra2005cavity}:
\begin{eqnarray}
\label{eq:kappa_coupling}
\kappa_{c}(\omega) ={-i} \sqrt{\frac{\Gamma_{c}}{2\pi}}e^{-i\frac{\omega}{c}L}\text{sinc}{\Big( (\omega-\omega_{c})\frac{L}{c}\Big)}.
\end{eqnarray}
{To derive this function, as it is demonstrated in \cite{dutra2005cavity}, one should first derive the modes corresponding to Maxwell's equations in a one-dimensional semi-infinite space incorporating a cavity made of a partially transparent single-layer mirror with negligible mirror thickness. Then, one can consider a model where the actual cavity is replaced by a perfect one which is then coupled to the semi-infinite reservoir delimited by the perfect cavity. By equating the inside modes of the partially transparent cavity obtained from Maxwell's equations to the discrete perfect cavity modes, and doing the same for the corresponding modes describing the outside, the expression in~\eqref{eq:kappa_coupling} can be obtained for sufficiently small transmission. We highlight that in this derivation there is no emitter initially considered in the system, and the coupling function $\kappa_{c}(\omega)$ describes the coupling of the empty cavity to the environment.}
 
 One can notice that the derived {inside-outside} representation does not feature {a constant cavity - reservoir coupling} as it is generally \textcolor{black}{assumed in standard derivation under certain conditions~\cite{gardiner00}. The standard approach, albeit leading to physically accurate results, can lead to mathematical inconsistencies (see \textcolor{black}{Appendix~\ref{app:delta_ev}}). Here, however, the cavity-reservoir coupling has a specific {form~\eqref{eq:kappa_coupling}}, which is obtained under mathematically explicitly defined conditions~\cite{dutra2005cavity} and can be treated straightforwardly.}
 
 \subsubsection{\label{sec:pseudo} {Pseudo-mode representation}}
 
{  We can define a pseudo-mode representation via ~\cite{multilayer}:
\begin{subequations}
\label{eq:effective_Hamiltonian}
\begin{eqnarray}
&&\hat{H}(t)= {H}_{A}(t)+\hat{H}_{AC}+\hat{H}_{C}\\
&&\hat{H}_{AC}=\hbar g \big(a^{\dagger}\sigma +\sigma^{\dagger }a\big),\\
&&\hat{H}_C=\hbar\left(\Delta-\Delta_{c}-i\frac{\Gamma_{c}}{2} \right)a^{\dagger}a,
\end{eqnarray}
\end{subequations}
where, for a single mode, 
\begin{subequations}
\begin{eqnarray}
a^{\dagger} \lvert \emptyset  \rangle & =&\lvert 1\rangle.
\end{eqnarray}
\end{subequations}
This representation is derived directly from the true-mode picture with the approximate coupling~\eqref{eq:eta_approx}. In general, the cavity mode $c$ in~\eqref{Hamiltonian} and $a$ in~\eqref{eq:effective_Hamiltonian} are different. $c$ is the perfect cavity mode, while $a$ is defined as~\cite{multilayer}: 
\begin{eqnarray}
\label{eq:pseudeo_mode_a_m}
a = \frac{1}{g}\int d\omega ~\hat{\eta} (\omega) \hat{a}_{\omega},
\end{eqnarray}
where $\hat{a}_{\omega}$ is the annihilation operator of the mode defined for the approximate coupling $\hat{\eta}(\omega)$.\\
 We highlight that Hamiltonian~\eqref{eq:effective_Hamiltonian} is for the open system  $\mathcal{A}\oplus\mathcal{C}$, where the reservoir is eliminated, while Hamiltonians~\eqref{Hamiltonian_universal} and~\eqref{Hamiltonian} both describe closed systems. Here, the annihilation operators $a$ and $c$ represent physically the same approximate modes but derived differently.
In the following we analyze these different representations by comparing the dynamics obtained via each Hamiltonian.
}

  \subsubsection{{Comparison and validity of the different representations}}
 
{ We numerically analyze the validity of different representations described in Fig.~\ref{fig:mode_representations}. We consider a single atom trapped in the cavity and assume an initial condition with zero photon. To differentiate the parameters of different representations, we denote the quantities corresponding to the true mode picture and the pseudo-mode picture with a tilde and a hat, respectively.
We commence by deriving the dynamics corresponding to the true mode representation. Here, one can denote the basis as $|i\rangle |\alpha\rangle\equiv |i,\alpha\rangle$, with $i$ labelling the atomic states and $\alpha$ describing the state of the continuum. The state in this basis can then be given by the following wavefunction:}
\begin{align}
\label{eq:true_mode_state}
\lvert \tilde{\psi} \rangle = \tilde{c}_{g,0}(t){\lvert g,\emptyset \rangle}+\tilde{c}_{e,0}(t){\lvert e,\emptyset \rangle} +\hspace{-0.3em}\int_0^{+\infty}\hspace{-1em}d\omega\,\tilde{c}_{f,1}(\omega,t)\lvert f,\mathbf{1_{\omega}} \rangle, 
\end{align}
with 
\begin{subequations}
\begin{eqnarray}
a^{\dagger}_{\omega}\mathbf{ \lvert \emptyset  \rangle} & =&\mathbf{ \lvert 1_{\omega} \rangle},\\
a_{\omega} \mathbf{\lvert 1_{\omega'} \rangle} & =& \delta(\omega-\omega'){\lvert \emptyset  \rangle}.
\end{eqnarray}
\end{subequations}
Using Hamiltonian \eqref{Hamiltonian_universal} in the time-dependent Schr\"odinger equation {with state~\eqref{eq:true_mode_state}},  we obtain the following dynamical equations:
\begin{subequations}
\label{eq:dynamics_full}
\begin{align}
i\dot{\tilde{c}}_{g,0}(t)& = \Omega \, \tilde{c}_{e,0}(t),\\
i\dot{\tilde{c}}_{e,0}(t)& = \Delta \, \tilde{c}_{e,0}(t)+\Omega \, \tilde{c}_{g,0}(t)\\ \nonumber
&{}+ \, i\int_0^{+\infty}\hspace{-1em}d\omega \; \eta(\omega) \, \tilde{c}_{f,1}(\omega,t),\\
i\dot{\tilde{c}}_{f,1}(\omega,t)& = ({\Delta-\Delta_{c}}+ \omega - \omega_{c})\tilde{c}_{f,1}(\omega,t)-i \eta^{\ast}(\omega) \tilde{c}_{e,0}(t),
\end{align}
\end{subequations}
{where $\eta(\omega)$ is the actual coupling function~\eqref{eq:eta_real}. If we take into account the approximation in~\eqref{eq:eta_approx}, similar equations to~\eqref{eq:dynamics_full} can be obtained for a state given by 
\begin{align}
\lvert \hat{\psi} \rangle = \hat{c}_{g,0}(t){\lvert g,\emptyset \rangle}+\hat{c}_{e,0}(t){\lvert e,\emptyset \rangle} +\hspace{-0.3em}\int_0^{+\infty}\hspace{-1em}d\omega\,\hat{c}_{f,1}(\omega,t)\lvert f,{1_{\omega}} \rangle,
\end{align}
with $\lvert 1_{\omega} \rangle = \hat{a}^{\dagger}_{\omega} \lvert \emptyset \rangle$.
Here, we can define the photon state of the cavity as~\cite{multilayer}:
\begin{eqnarray}
\label{eq:out_photon_cav}
\lvert 1\rangle & =& \frac{1}{g}\int_0^{+\infty}\hspace{-1em} d\omega~ \hat{\eta}^{\ast}(\omega) \hat{a}^{\dagger}_{\omega} \lvert \emptyset  \rangle,
\end{eqnarray}
which follows from the definition~\eqref{eq:pseudeo_mode_a_m}. Consequently, by using Hamiltonian~\eqref{eq:effective_Hamiltonian} we get the dynamics corresponding to the pseudo-mode representation, on a reduced basis $\{\lvert g,\emptyset \rangle, \lvert e,\emptyset \rangle, \lvert f,1 \rangle \}$, for the state given by 
\begin{eqnarray*}
\lvert \psi_{\rm{eff}} \rangle = \hat{c}_{g,0}(t){\lvert g,\emptyset \rangle}+\hat{c}_{e,0}(t){\lvert e,\emptyset \rangle} +\hat{c}_{f,1}(t)\lvert f,1 \rangle :
\end{eqnarray*}
\begin{subequations}
\label{eq:dynamics_effective}
\begin{eqnarray}
i\dot{\hat{c}}_{g,0}(t)& =& \Omega \, \hat{c}_{e,0}(t),\\
i\dot{\hat{c}}_{e,0}(t)& =& \Delta \, \hat{c}_{e,0}(t)+\Omega \, \hat{c}_{g,0}(t)+g \,\hat{c}_{f,1}(t)\\ 
i\dot{\hat{c}}_{f,1}(t)& = &\left(\Delta-\Delta_{c}-i\frac{\Gamma_{c}}{2}\right)\hat{c}_{f,1}(t)+g\, \hat{c}_{e,0}(t).
\end{eqnarray}
\end{subequations}
}

Unlike the case of {true mode picture}, in the inside-outside representation there is a separation of the photon state into inside and the outside ones. {Thus, the basis splits into {the following} states:  $\{|g,\emptyset \rangle,|e,\emptyset \rangle, |f,{1_{\text{in}}},\emptyset_{\text{out}}\rangle, |f,\emptyset_{\text{in}},1_{\omega ,\text{out}}\rangle\}$, where the indices \textit{in} and \textit{out} indicate the photon state inside and the outside of the cavity respectively,  and 
 \begin{subequations}
 \begin{eqnarray}
 c^{\dagger}  \lvert \emptyset \rangle & =& \lvert 1_{\text{in}} \rangle ,\\
 c  \lvert 1_{\text{in}} \rangle  & =&  \lvert \emptyset \rangle,\\
b^{\dagger}_{\omega} \lvert \emptyset \rangle & =& \lvert 1_{\omega, \text{out}} \rangle ,\\
b_{\omega} \lvert 1_{\omega', \text{out}} \rangle & =& \delta(\omega-\omega')\lvert \emptyset \rangle.
\end{eqnarray}
\end{subequations}}\color{black}
The dynamical equations corresponding to the Hamiltonian \eqref{Hamiltonian} with the state 
\begin{eqnarray}
\nonumber
\lvert \psi \rangle&=& {c_{g,0}(t)\lvert g,\emptyset \rangle+c_{e,0}(t)\lvert e,\emptyset \rangle }+c_{f,1,0}(t)\lvert f,1_{\text{in}},\emptyset_{\text{out}} \rangle \\ \label{eq:state_in_out}
&{}& + \int_0^{+\infty} \hspace{-1.2em}  d\omega \, c_{f,0,1}(\omega,t)\lvert f,\emptyset_{\text{in}}, {1_{\omega, \text{out}}} \rangle
\end{eqnarray}
become: 
\begin{subequations}
\label{eq:dynamics_in_out}
\begin{align}
i{\dot{c}_{g,0}}(t) &= \Omega \,{c_{e,0}}(t),\\
i{\dot{c}_{e,0}}(t) &= \Delta \, {c_{e,0}}(t)+\Omega \, {c_{g,0}}(t)+{g \, c_{f,1,0}(t)},\\ \nonumber
i\dot{c}_{f,1,0}(t) &= {(\Delta-\Delta_{c})} c_{f,1,0}+g \, {c_{e,0}}(t)\\
&{}-i\int_0^{+\infty}\hspace{-1em}d\omega \; \kappa_{{c}}^{\ast}(\omega) \, c_{f,0,1}(\omega,t),\\
i\dot{c}_{f,0,1}(\omega,t) &= \left( \Delta-\Delta_{c}+\omega - \omega_{c}\right)c_{f,0,1}(\omega,t)\\ \nonumber
&{}+i \kappa_{c}(\omega) \, c_{f,1,0}(t).
\end{align}
\end{subequations}

\begin{figure*}[!ht]
\begin{center}
\includegraphics[scale=0.48]{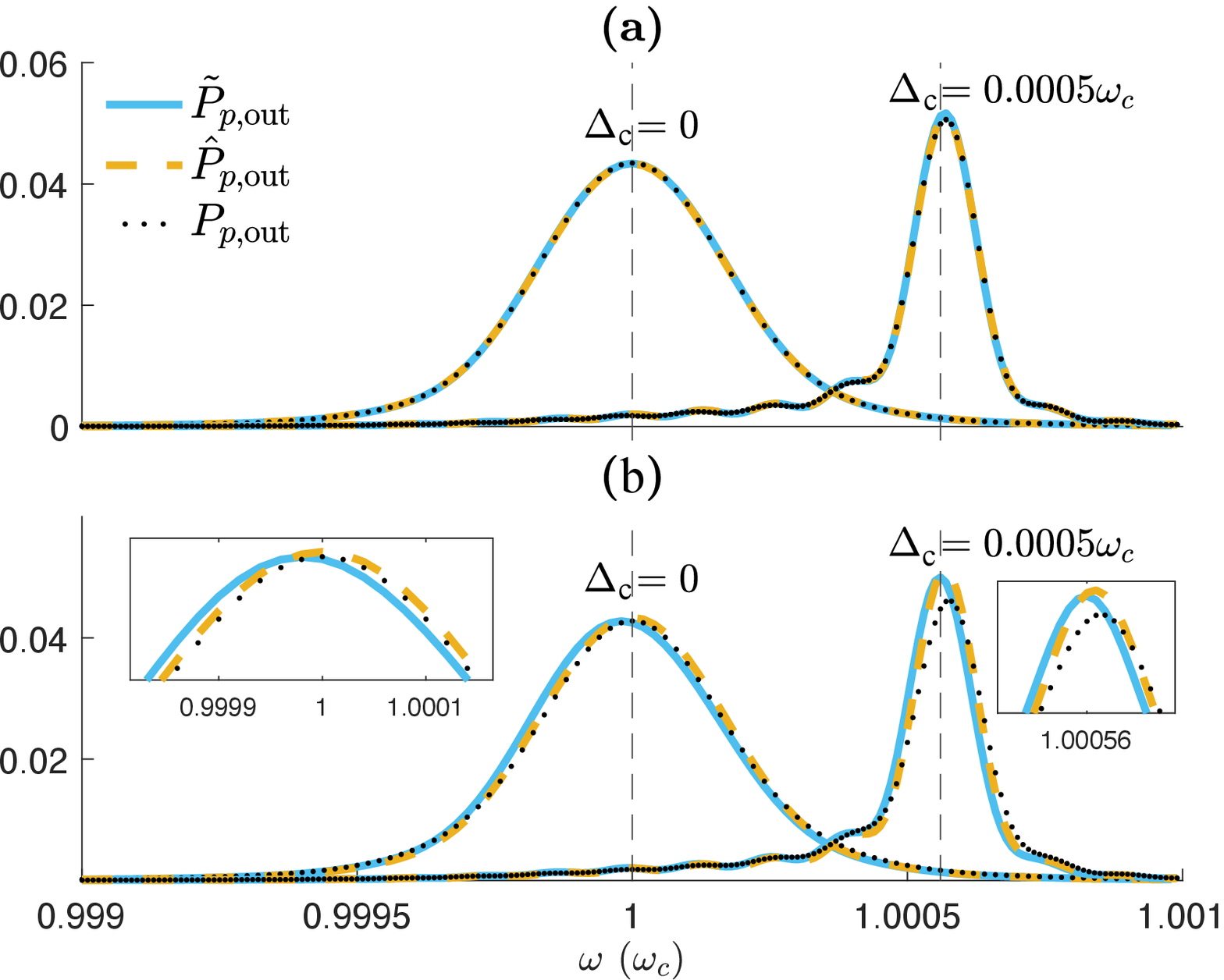}
\includegraphics[scale=0.48]{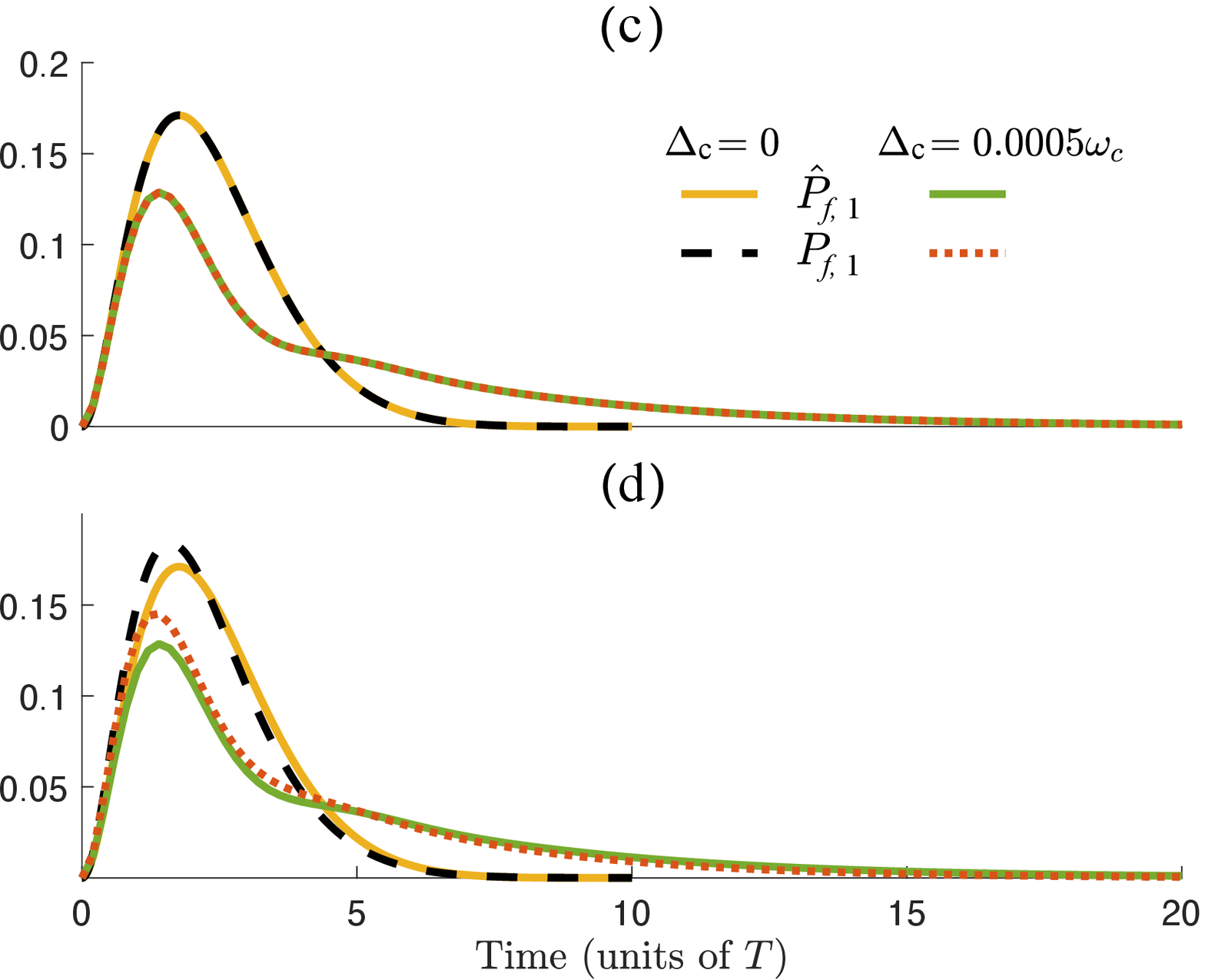}
\caption{{Comparison of the single photon shape obtained via true mode, inside-outside and pseudo-mode pictures. \textcolor{black}{Left panel}: Spectral shape of outgoing photon with  $(|g|, \Gamma_{c}, \omega_{c}) \times T=(0.6, 2, 2416)$, where $\omega_{c} = m\frac{\pi c}{L}$, $m$ being the number of antinodes inside the cavity. Each curve corresponds to the different representations: $\tilde{P}_{p,\rm{out}}=\lvert\mathbf{\tilde{c}}_{f,1}(\omega,t_{\rm{f}})\rvert^{2}$ (true-mode picture), $\hat{P}_{p,\rm{out}}=\lvert\mathbf{\hat{c}}_{f,1}(\omega,t_{\rm{f}})\rvert^{2}$ (pseudo-mode picture), $P_{p,\rm{out}}=\lvert\mathbf{c}_{f,0,1}(\omega,t_{\rm{f}})\rvert^{2}$ (inside-outside picture) (see the definition of the dimensionless parameters $\mathbf{c}_{i}$ in Appendix~\ref{app:integration}), where $t_{\rm{f}}$ is the final time when the photon is in its steady state. \textcolor{black}{(a)}: $t_{f}/T=10$, $\Gamma_{c}=2/T$ is obtained for a cavity with a length $L=L_{0}$ and a fictitious refractive index $n = 27.735$, leading to the reflectivity $R=|\underline{r}|^{2}=e^{-2L\Gamma_{c}/c}\approx 0.995$ (with $\frac{L_{0}}{cT}=0.0013$), where $L_{0}$ is the length for which the cavity sustains single fundamental mode, i.e., $m=1$. \textcolor{black}{(b)}: $t_{f}/T=20$, $L=165L_{0}$ and $n = 2.1756$, leading to the same $\Gamma_{c}= 2/T$, for the $m=165th$ mode, with the reflectivity $R\approx 0.42$. \\ 
\textcolor{black}{Right panel}: Time profile of the outgoing photon, with \textcolor{black}{the figures (c) and (d) obtained via the same parameters introduced in (a) and (b), respectively}. $\hat{P}_{f,1}=\lvert \hat{c}_{f,1}(t)\rvert^{2}$, $P_{f,1}=\lvert c_{f,1,0}(t)\rvert^{2}$ are the photon states derived from the pseudo-mode and inside-outside representations, respectively.}}
 \label{equivalence}
 \end{center}
\end{figure*}

{
In order to examine the validity limits of the approximate models derived above, we compare the dynamics via solving Eqs.~\eqref{eq:dynamics_full},~\eqref{eq:dynamics_effective} and~\eqref{eq:dynamics_in_out} (see the details of integration of eqs.~\eqref{eq:dynamics_full} and~\eqref{eq:dynamics_in_out} in Appendix~\ref{app:integration}). For this analysis, the way the atom is driven to its excited state is irrelevant. Thus, we assume that the atom is initially in the excited state $\lvert e \rangle$ and there is no laser field applied, i.e. $\Omega=0$. We analyze a regime where there are no Rabi oscillations between the atom and the produced photon, i.e., the leakage from the cavity is stronger than the atom-cavity coupling: $\Gamma_{c}>g$.  
Both the effective Hamiltonian as well as the inside-outside representation are derived under the assumption of having a high-$Q$ cavity, i.e., $\Gamma_{c}\ll \omega_{c}$. In Fig.~\ref{equivalence}, we present the results obtained via different representations for a cavity with a fixed quality factor: ${\omega_{c}}/{\Gamma_{c}}\approx 1200$. This factor is obtained  either by fixing the mirror refractive index and changing the length of the cavity or vice versa. In Fig.~\ref{equivalence}(a), the cavity length is such that it sustains half a wavelength of cavity resonance wavelength $\lambda_{c}$: $L=L_{0}=\lambda_{c}/2$. Therefore, there is only a single mode that the atom can couple to, making the cavity finesse the same as the quality factor: $\Delta_{\omega}/\Gamma_{c}\approx 1200$. As we can see from the figure, both for detuned and non-detuned cases, the photons obtained with pseudo-mode and inside-outside representations match the true-mode representation obtained from eq.~\eqref{eq:dynamics_full}, with the coupling~\eqref{eq:eta_real} \footnote{In the simulation we use the actual response function $\eqref{eq:app:actual_resp}$ derived for a single layer mirror with a thickness $\delta = \lambda_{c}/(4n)$, and the atom is placed at $x_{a}=-L/2$ }. Furthermore, in Fig.~\ref{equivalence}(b), we consider a cavity of longer length and a mirror of lower refractive index. To have the same atom-cavity coupling rate, we change the value of the dipole moment of the atom. While having the same quality factor, here we get a cavity finesse $\Delta_{\omega}/\Gamma_{c}\approx 7$. This low finesse makes the transition from eq.~\eqref{eq:eta_real} to eq.~\eqref{eq:coupling_full_Lor} less accurate, i.e., the Lorentzians corresponding to each mode are not separated well enough to consider $\sqrt{\sum f_{m}}\approx \sum {\sqrt{f_{m}}}$. We emphasise that the approximation \eqref{eq:coupling_full_Lor} is used in the derivation of the pseudo-mode picture as well as the inside-outside representation. Hence, it leads to a mismatch between the approximate and the actual representations. Equivalently, as it is shown in~\cite{multilayer}, in order for the pseudo-mode derivation to work, the following condition should hold: $\left(\frac{\Gamma_{c}}{c}(x_{A}+L)\right)^{2}\ll1$. Evidently, when we increase the cavity length while keeping $\Gamma_{c}$ the same, this condition is not well satisfied, breaking the validity of this representation.  On the other hand, as mentioned before, the inside-outside representation is derived for the cavities with low transmission rate, given by $|\underline{t}|^{2}=1-e^{-2L\Gamma_{c}/c}$ (see Appendix~\ref{app:Lorentzian_separation}). For this long cavity scenario, similar to the previous argument, this term fails to satisfy the condition $|\underline{t}|^{2}\ll 1$ ($|\underline{t}|^{2}\approx 0.58$), which leads to the mismatch between the inside-outside and the true-mode representations. Finally, we can notice that even in the case of $\Delta_{c}=0$, the photon obtained from the actual model is slightly shifted from the resonance frequency $\omega_{c}$. This is because in the actual model where we use the response function $T(\omega)$, apart from the fundamental mode $\omega_{c}$, there are other modes, which, combined with the low finesse of the cavity, affect the produced photon.}

{
In Fig.~\ref{equivalence}\textcolor{black}{(c) and (d)}, we study the corresponding photon shape in the time domain. Here we compare the cavity photon state obtained from the pseudo-mode ($\hat{c}_{f,1}(t)$) and the inside-outside representations ($c_{f,1,0}(t)$). As we can see from Eq.~\eqref{eq:dynamics_effective}, the photon state $\hat{c}_{f,1}(t)$ depends only on the parameters $\Gamma_{c}$ and $g$, and these parameters are fixed, hence $\hat{P}_{f,1}$ is the same both for high (top figure) and low (bottom figure) finesse scenarios. On the other hand, in the inside-outside representation, the coupling $\kappa_{c}(\omega)$ explicitly depends on the cavity length, leading to different curves for the state $P_{f,1}$. This, combined with the arguments introduced in the analysis of Fig.~\ref{equivalence}(a) \textcolor{black}{and (b)}, lead to the differences between the photon obtained via pseudo-mode and inside-outside representations.}

 {In the following, we derive the well-known master equation starting from the inside-outside representation. We use a method different from the standard derivation obtained by phenomenological use of the pseudo-mode Hamiltonian.}

\subsection{Heisenberg-Langevin equations, Poynting vector, and photon fluxes}

We wish to derive the effective dynamics of the atom-cavity system {$\mathcal{S}=\mathcal{A}\oplus\mathcal{C}$}, coupled to the reservoir, {from first principles}. Our aim is to control the production of an outgoing photon leaking from the cavity by driving specifically the atom in the cavity by the external field. {We will use the convenient inside-outside representation as it will allow a clear identification and characterization of the leaking photon propagating in free space. Here, we study the case presented in Fig.~\ref{1atc}, where the atom is initially in the ground state, and we consider the two-photon resonance condition: $\Delta = \Delta_{c}$, leading to $\omega_{gf}=\omega_c-\omega_0$.} We use the Poynting vector that we {derive from the true mode representation} and define in the Heisenberg picture. {We highlight that one could use the Poynting vector derived in~\cite{Blow_1990} and generalize it to the situation with the presence of the cavity, however, as we show below, the Poynting vector that we derive from first principles is different from this one. We then derive} the effective model in two steps: we first define an outgoing flux of photon which is connected to the quantum average of the Heisenberg evolution of the cavity operator $c^{\dagger}c$. Next we derive a master equation of the system $\mathcal{S}$ by eliminating the reservoir degrees of freedom, which will allow the calculation of the quantum averages.

\subsubsection{Equations of motion for the operators}

\begin{figure}[!h]
\begin{center}
\includegraphics[scale=.6]{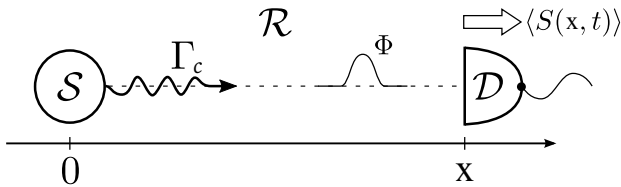}
\caption{\small {Sketch of the photodetection: the source system $\mathcal{S}$ emits a photon with decay rate $\Gamma_{c}$ at position 0, towards a detector $\mathcal{D}$ at a position $x$ through the reservoir $\mathcal{R}$. The photon flux $\Phi$ is measured using the data on the averaged quantized Poynting vector $\langle{S}(x,t)\rangle$.}}
\label{detection}
\end{center}
\end{figure}

First, we derive the equations of motion in the Heisenberg picture for
{the          reservoir          operator         $b_{\omega}(t)\equiv
U^{\dagger}(t,t_0)b_{\omega}U(t,t_0)$  with  $U(t,t_0)$} being
the propagator of the total Hamiltonian ${H}(t)$, whose Heisenberg
picture                                                    reads
${H}^{(H)}(t)=U^{\dagger}(t,t_0){H}(t)U(t,t_0)$.
From $\dot{\mathcal{O}}=-\frac{i}{\hbar}[\mathcal{O}(t),{H}^{(H)}(t)]$ for an operator $\mathcal{O}$, assumed time-independent in the Schr\"odinger picture, and written as $\mathcal{O}^{(H)}(t)\equiv\mathcal{O}(t)=U^{\dagger}(t,t_0)\mathcal{O}U(t,t_0)$ in the Heisenberg picture,
we write the Heisenberg-Langevin equations:
\begin{subequations}
\label{Heis-Lang2full}
\begin{align}
\label{Heis-Lang2}
\dot{b}_{\omega}(t)&=-i\omega b_{\omega}(t)+\kappa_{c}(\omega)c(t),\\
\dot{c}(t)&=-i\omega_c c(t)- \int_0^{+\infty}\hspace{-1em} d\omega~\kappa_{{c}}^{\ast}(\omega)b_{\omega}(t)-ig\sigma(t).
\end{align}
\end{subequations}
{In the following, we omit the $(H)$ superscript for the Heisenberg picture Hamiltonian ${H}^{(H)}(t) \equiv {H}(t)$}.  The energy
carried by the photons leaking from the cavity can be characterized by
the  Poynting   vector  operator  in   the  Heisenberg  picture. {We derive the Poynting vector, using the electromagnetic fields outside the cavity in the true mode representation \cite{dutra2005cavity,vogel2006quantum,multilayer} (see the details of the derivation in Appendix~\ref{app:poynting}):
\begin{align}
S(x,t)=\frac{\hbar}{2\pi \mathcal{A}}\int^{\infty}_{0} \hspace{-1em}d\omega d\omega' \sqrt{\omega\omega'}R_{\omega}R^{\ast}_{\omega'}e^{i(\omega-\omega')\frac{x}{c}}a_{\omega'}^{\dagger}(t)a_{\omega}(t),
\end{align}
where the expression of $R_{\omega}$ is given in Appendix~\ref{app:poynting}.
 Via moving from the true mode representation to the inside-outside representation, we get the following Poynting vector (Appendix~\ref{app:poynting}):
\begin{eqnarray}
\label{eq:Poynting_out}
S(x,t) = \frac{\hbar \omega_{c}}{\mathcal{A}}b^{\dagger}(x,t)b(x,t),
\end{eqnarray}
where we have introduced the integrated reservoir operator 
\begin{eqnarray}
\label{eq:in_bath_op}
b(x,t) :=  \frac{1}{\sqrt{\Gamma_{c}}}\int^{\infty}_{0} \hspace{-1em}d\omega \kappa_{c}^{\ast}(\omega) e^{i\frac{\omega}{c}x}b_{\omega}(t). 
\end{eqnarray}
Taking this into account, we can define a single photon state propagating from the cavity via leakage $\Gamma_{c}\;(\kappa_{c}(\omega))$ and defined for $x>0$:
\begin{eqnarray}
\label{eq:out_photon}
\lvert 1_{\rm{out}}(x,t) \rangle = b^{\dagger}(x>0,t) \lvert \emptyset \rangle ,
\end{eqnarray} 
where} we have assumed a propagation  with increasing $x$
and the cavity emitter at position $x=0$ (see Fig. \ref{detection}). We emphasize that the time dependence arises only from the Heisenberg picture of the reservoir operator $b_{\omega}$.

\subsubsection{Integrated reservoir operators - Input output relation}

{In eq.~\eqref{eq:in_bath_op} we have defined the integrated reservoir operator, which we calculate by integrating \eqref{Heis-Lang2} from an initial time $t_0$ to $t$:}
\begin{subequations}
\begin{align}
\label{bdef}
& b(x,t):={\frac{1}{\sqrt{\Gamma_{c}}}}\int_0^{+\infty}\hspace{-1em} d\omega~ \kappa_{{c}}^{\ast}(\omega) b_{\omega}(t)e^{i\omega\frac{x}{c}}\\
\label{bdef_b}
&=b_{\text{in}}\Bigl(t-\frac{x}{c}\Bigr)+\int_{t_0}^t \hspace{-0.5em} dt'  \hspace{-0.3em} \int_0^{+\infty}\hspace{-1em} d\omega \frac{|\kappa_{c}(\omega)|^{2}}{{\sqrt{\Gamma_{c}}}}c(t')e^{-i\omega(t-t')}e^{i\omega \frac{x}{c}}
\end{align}
\end{subequations}
with the \textit{input} operator defined similarly:
\begin{equation}
b_{\text{in}}\Bigl(t-\frac{x}{c}\Bigr):= {\frac{1}{\sqrt{\Gamma_{c}}}}\int_0^{+\infty}\hspace{-1em} d\omega~ \kappa_{c}^{\ast}(\omega)b_{\omega}(t_{0}) e^{-i\omega(t-t_0-\frac{x}{c})}.
\end{equation}
{One can notice, that the definition~\eqref{bdef} is different from the standard definition, where $b(x,t)$ is defined via a Fourier transform of $b_{\omega}$ \cite{gardiner00,Graham1968,Raymer043819} \textcolor{black}{and a flat continuum (see Appendix~\ref{app:delta_ev})}. Instead, here we have the natural introduction of the function $\kappa_{c}(\omega)$ in the definition, and its explicit form allows one to straightforwardly derive the equation for output operator. }

In order to evaluate the integral over the frequency in~\eqref{bdef_b} we use the expression \eqref{eq:kappa_coupling}.
This gives for the integrated reservoir operator (see the details in Appendix \ref{app:evaluation_of_integral}):
\begin{equation}
\label{bath}
 b(x,t)=b_{\text{in}}\Bigl(t-\frac{x}{c}\Bigr)+{\sqrt{\Gamma_c}} c\Bigl(t-\frac{x}{c}\Bigr),
\end{equation}
where we have assumed $t\gg \frac{2L}{c}$, $t>t_{0}+\frac{x}{c}+\frac{2L}{c}$.
We further neglect the cavity length, assuming that the traveling distance of interest is much larger than the cavity length. Also, considering that the dynamics evolves in much larger times than the round trip time $\frac{2L}{c}$ of the produced photon (coarse-grained approximation), we get: $t\gg 0$, $t>t_{0}+\frac{x}{c}$ and $x>0$ . We define the \textit{output} operator
\begin{equation}
\label{output_def}
b_{\text{out}}(t-x/c):=  b(x>0,t);
\end{equation}
hence,
\begin{equation}
\label{inout}
b_{\text{out}}\Bigl(t-\frac{x}{c}\Bigr)=b_{\text{in}}\Bigl(t-\frac{x}{c}\Bigr)
+{\sqrt{\Gamma_c}}\ c\Bigl(t-\frac{x}{c}\Bigr),
\end{equation}
which  is  recognized  as  the  input-output  relation  \cite{gardiner00}.  
We highlight that the input-output relation here is a consequence of the concrete form of $\kappa_{c}(\omega)$ [Eq.~\eqref{eq:kappa_coupling}], which justifies thus the Markov approximation. This way of formulation allows
direct and transparent interpretation of the $b_{\text{out}}$ operator
through    the     Poynting    vector    as     shown    below    [see
  Eq. \eqref{flux_bout}].
  
 At the cavity position, $x=0$, we obtain the integrated reservoir operator (see Appendix \ref{app:evaluation_of_integral}):
\begin{equation}
\label{bath0}
b_0(t)\equiv  b(x=0,t)=b_{\text{in}}(t)+\frac{{\sqrt{\Gamma_c}}}{2}
\ c(t).
\end{equation}
This expression \eqref{bath0} is used in the next subsection to derive the master equation in the cavity.

We can also simplify the Heisengerg-Langevin equation for $c(t)$ as:
\begin{equation}
\label{eqc}
\dot{c}(t)=-\left(i\omega_c+\frac{\Gamma_c}{2}\right) c(t)-{\sqrt{\Gamma_{c}}}b_{\text{in}}(t)-ig \, \sigma (t).
 \end{equation}

\subsubsection{{Photon flux}}
\label{Poynting_sub}
{Using the results obtained in the previous section, we can write the Poynting vector~\eqref{eq:Poynting_out} in the inside-outside representation as follows:}
\begin{equation}
{S}(x>0,t)=\frac{\hbar\omega_c}{\mathcal{A}}b^{\dagger}\Bigl(t-\frac{x}{c}\Bigr) b\Bigl(t-\frac{x}{c}\Bigr).
\end{equation}
For  a given state  (or density  matrix), the  amount of  energy going
through the  field mode area  $\mathcal{A}$, during the time  $dt$, is
the quantum  average of the flux  of the Poynting  vector through this
area:  $\mathcal{A}\langle{S}(x,t)\rangle dt=\hbar\omega_c \langle
    b^{\dagger}(x,t)    b(x,t)\rangle    dt$.    Normalizing    by
$\hbar\omega_c$,   we    get   the   averaged    number   of   photons
$dn(x,t)\equiv\langle   b^{\dagger}(x,t)  b(x,t)\rangle dt$  going
through the mode  area during $dt$, defining the  photon flux (written
here for $x>0$):
\begin{equation}
\label{flux_bout}
\Phi(x,t):= \frac{d n(x,t)}{dt}
=\Bigl \langle b^{\dagger}\Bigl(t-\frac{x}{c}\Bigr)b\Bigl(t-\frac{x}{c}\Bigr)\Bigr\rangle.
\end{equation}
Recalling that $b(t-x/c)$ is the output operator \eqref{output_def}, we emphasize that this relation gives the connection between the photon flux and this output operator.

If we choose the state of the reservoir to be initially a vacuum state: $\rho(t_0)=\rho_S(t_0)\otimes |\emptyset \rangle \langle \emptyset |$, the average of the terms involving $b_{\text{in}},b_{\text{in}}^{\dagger}$ in the expression of the flux nullifies. This gives the expression of the outgoing photon flux through the semi-transparent mirror for $t>t_0+\frac{x}{c}, {x}>0$:
\begin{equation}
\label{flux}
{\Phi(x,t)=\Gamma_c\Bigl\langle c^{\dagger}\Bigl(t-\frac{x}{c}\Bigr)c\Bigl(t-\frac{x}{c}\Bigr)\Bigr\rangle.}
\end{equation}
This key result shows that one can determine the flux from the quantum average of the dynamics of the cavity photon number in the Heisenberg picture \cite{Gogyan_A}.

In the following subsection, we derive the effective master equation reduced to the system $\mathcal{S}$ which is used to calculate the quantum average of \eqref{flux_bout} in order to derive the flux.

\subsection{The master equation for the system dynamics}

The system dynamics from the above inside-outside representation can be characterized by a master equation which is shown to be of Lindblad form. We follow the derivation of \cite{Louisell,Carmichael,gardiner00,Breuer}. We need first to derive the Heisenberg equation of motion of the operators $X_S(t)=U^{\dagger}(t,t_0)X_SU(t,t_0)$ of the system in the Heisenberg picture.

The dynamics of $X_S(t)$ is determined from the Heisenberg\textcolor{black}{-Langevin} equation (see the details of the following calculation in Appendix \ref{app:master_eq}):
\begin{eqnarray}
\frac{d}{dt} X_S(t)=-\frac{i}{\hbar}\Bigl[X_S(t),{H}_S^{(H)}(t)\Bigr]+\mathcal{D}^{\dagger}_{\text{in},t}
\bigl(X_S(t)\bigr)&&\nonumber\\ 
\label{eq:X_S_dynamics}
  +\, \Gamma_c\bigl(c^{\dagger}(t)X_S(t)c(t) -\textstyle\frac{1}{2}\displaystyle\{c^{\dagger}(t)c(t),X_S(t)\}\bigr)&&, \quad
\label{X}
\end{eqnarray}
where $\{A,B\}=AB+BA$ denotes the anticommutation relation, $\mathcal{D}^{\dagger}_{\text{in},t}(\cdot)$ is a time-dependent dissipator part involving $b_{\text{in}}(t)$, acting on $X_S(t)$, and ${H}_S^{(H)}(t)=U^{\dagger}(t,t_0){H}_S(t)U(t,t_0)$, {with $H_{S}$ being the system Hamiltonian:
\begin{eqnarray}
\label{eq:HSyst}
H_{S}(t)=H_{A}+H_{AC}+H_{C}.
\end{eqnarray}
}
In Eq.~\eqref{eq:X_S_dynamics} we have used the reservoir integrated operator \eqref{bath0} at the position $x=0$ of the cavity.

We define the expectation value of $X_S$:
\begin{equation}
\label{Tr}
\langle X_S(t)\rangle =\mathrm{Tr}_S\{X_S\rho_S(t)\}=\mathrm{Tr}\{X_S(t)\rho(t_0)\},
\end{equation}
where $\rho(t_0)=\rho_S(t_0)\otimes\rho_R(t_0)$ is the complete density operator and $\rho_S(t)=\mathrm{Tr}_R\{U(t,t_0)\rho(t_0)U^{\dagger}(t,t_0)\}$ is the reduced density operator describing $\mathcal{S}$ with partial trace $\mathrm{Tr}_{R}\{\cdot\}$ eliminating the degrees of freedom corresponding to its subscript.

We here assume that the reservoir is initially a vacuum state $\rho_R(t_0)\equiv|\emptyset \rangle\langle\emptyset|$ such that $\mathcal{D}^{\dagger}_{\text{in},t}(\cdot)$ cancels out in averaging. Finally, averaging Eq. \eqref{X}, we find the master equation of Lindblad form for $\rho_S(t)$:
\begin{eqnarray}
\label{eq:master_eq}
\frac{d}{dt}\rho_S(t)&=&-\frac{i}{\hbar}[{H}_S(t),\rho_S(t)]\nonumber\\
&{}&+\,\Gamma_c\big(c\rho_S(t)c^{\dagger} -\textstyle\frac{1}{2}\displaystyle\{c^{\dagger}c,\rho_S(t)\}\big),
\label{rho_S}
\end{eqnarray}
where, here, all system operators $\sigma,c$ are time-independent (Schr\"odinger picture), and the remaining time-dependence of ${H}_S(t)$ is due to the driving field $\Omega (t)$.


\section{\label{sec:prod_sing} Production of a single photon by a driven atom trapped in cavity}

As an application, we derive from the preceding analysis a model for generation of a single photon using a leaking cavity containing one atom driven by a pulsed laser of Rabi frequency $\Omega(t)$. The production of a single photon in such a system has been demonstrated with
an atom flying through the cavity in a resonant stimulated Raman adiabatic passage configuration \cite{Vasilev_2010,Nisbet_Jones_2011,Kuhn1999} and for a trapped ion in a cavity \cite{Keller_njp}. We next show that a large cavity detuning and an effective weak coupling regime allow the direct and simple control of the photon shape.

\subsection{The model}

Since the system of interest involves only the atom and the cavity, in the effective model, 
the basis {introduced in the inside-outside representation} reduces to $\{|g,\emptyset \rangle,|e,\emptyset \rangle,|f,1\rangle,|f,\emptyset \rangle\}$  (see Fig.~\ref{1atc}), where the third label is dropped due to the elimination of the reservoir degrees of freedom and the labeling ``in'' is omitted. Such dynamics involves the Lindblad equation derived previously (we omit the subscript $S$ for $\rho$):
\begin{equation}
\frac{d}{dt}\rho(t)=-\frac{i}{{\hbar}}[H_S(t),\rho(t)]+\mathcal{L}[\rho(t)],\label{lindblad}
\end{equation}
with the dissipator $\mathcal{L}[\rho]=\Gamma_c(c\rho c^{\dagger} - \frac{1}{2}\{\rho, c^{\dagger}c\})$. Equation \eqref{lindblad} can be rewritten as
\begin{equation}
\frac{d}{dt}\rho(t)=-\frac{i}{{\hbar}}(\tilde{H}(t)\rho(t) -\rho(t)\tilde{H}^{\dagger}(t))+\Gamma_c\ c\rho(t) c^{\dagger},\label{lindblad2}
\end{equation}
where we introduced an anti-Hermitian dissipative Hamiltonian $\tilde{H}(t)=H_S(t)-i{\hbar}\frac{\Gamma_c}{2}c^{\dagger}c$, {equivalent to~\eqref{eq:effective_Hamiltonian}}. Expressing the Hamiltonian in a matrix form 
\begin{subequations}
\begin{eqnarray}
H_S(t)&=&\hbar\left[\begin{array}{cc}\mathbf{A}(t)& \left [0\right ]_{3\times 1}\\ \left [0\right ]_{1\times 3}& {-\omega_{c}}\end{array}\right], \label{hac}\\
\mathbf{A}(t)&=&\left[\begin{array}{ccc} 0&\Omega(t)&0\\ \Omega(t)&\Delta &g\\0&g&0\end{array}\right]
\end{eqnarray}
\end{subequations}
shows two  decoupled dynamical blocks $\mathbf{A}(t)$
and $\{-\omega_{c}\}$. From the density matrix
\begin{equation}
\rho(t)=\left[\begin{array}{cc}\rho_{\mathbf{AA}}(t)&\rho_{\mathbf{A}0}(t)\\
\rho_{0\mathbf{A}}(t)&\rho_{00}(t)\end{array}\right],
\end{equation}
we split Eq. \eqref{lindblad2} into two equations:
\begin{subequations}
\begin{eqnarray}
\dot{\rho}_{\mathbf{AA}}&=&-i(\mathbf{\tilde{A}}(t)\rho_{\mathbf{AA}}(t)-\rho_{\mathbf{AA}}(t)\mathbf{\tilde{A}}^{\dagger}(t)),
\label{schroda}\\
\dot{\rho}_{00}&=&\Gamma_c \mathbf{D}\rho_{\mathbf{AA}}(t)\mathbf{D}^{\dagger},\label{rhodelta}
\end{eqnarray}
\end{subequations}
where $\mathbf{D}=[0,0,1]$ is a block from the matrix representation $\mathbf{c}$ of the annihilation operator $c$, $\mathbf{\tilde{A}}(t)=\mathbf{A}(t)-\frac{i}{2}{\Gamma_c} \mathbf{D}^{\dagger}\mathbf{D}$. Choosing the initial condition in $|g,\emptyset \rangle$ makes the dynamics not involving $\rho_{A0}$ and Eq. \eqref{schroda} corresponds thus to a Schr\"odinger equation with losses (i.e. with a non-Hermitian Hamiltonian) {derived in Eq.~\eqref{eq:dynamics_effective}}, i.e. $\mathrm{Tr}\rho_{AA}<1$:
\begin{equation}
i\frac{\partial}{\partial t}|{\psi_{\rm{eff}}}\rangle=\left[\begin{array}{ccc} 0&\Omega(t)&0\\ \Omega(t)&\Delta &g\\0&g&-i\frac{\Gamma_c}{2}\end{array}\right]|{\psi_{\rm{eff}}}\rangle
\label{schroda_}
\end{equation}
with $\psi_{\rm{eff}}$ being the state in pseudo-mode picture. 
The population lost from the subspace spanned by the states $\{|g,\emptyset \rangle,|e,\emptyset \rangle,|f,1\rangle\}$ (on which the block $\mathbf{A}$ is defined) is collected in state $|f,\emptyset \rangle$ (on which the block ${\{-\omega_{c}\}}$ is defined), so that the whole system is closed: $P_{g,0}(t)+P_{e,0}(t)+P_{f,1}(t)+P_{f,0}(t)=1$ with            the population $P_{i,n}(t)=\langle i,n|\rho(t)|i,n\rangle=|c_{i,n}|^2$. {We highlight, that Eq.~\eqref{schroda_} is obtained from the inside-outside picture, for the cavity of mode $c$. This equation coincides with Eq.~\eqref{eq:dynamics_effective}, obtained from pseudo-mode picture, hence in this limit the modes $c$ and $a$ are the same.}

Rewriting \eqref{rhodelta} we get:
\begin{equation}
\frac{d}{dt} P_{f,0}(t)=\Gamma_c P_{f,1}(t).
\end{equation}
  On  the  other  hand,  from  the
definition of  the average $\langle\mathcal{O}\rangle=\mathrm{Tr}(\rho\mathcal{O})$,
one  can write  the  photon  flux \eqref{flux}  in  terms of  the
populations:
\begin{equation}
\Phi(t)\equiv\frac{d n(t)}{dt}=\Gamma_c P_{f,1}(t).\label{flx1}
\end{equation}
We can then  identify $P_{f,0}(t)$ as the number of the outgoing
photons:  $P_{f,0}(t)\equiv  n(t)$.   The  scheme enables  us  to
derive the  shape of  the leaking photon,  through its  flux $\Phi(t)$
from   the  atom-cavity   dynamics,   which  is   determined  by   the
Schr\"odinger equation \eqref{schroda_}.

\subsection{The scheme for a large detuning}

\begin{figure}[!ht]
\includegraphics[width=0.5\textwidth]{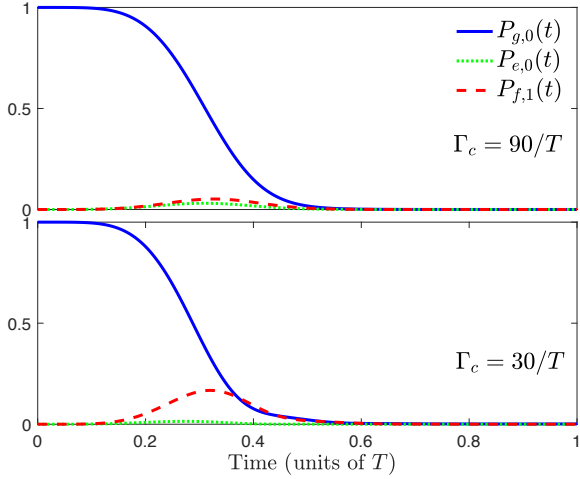}
\caption{Dynamics corresponding to Eq.~\eqref{schroda_} for cavities with different $\Gamma_{c}$ factors. The system parameters are: $\Omega(t)=\Omega_{0}\sin^{2}{\left(\frac{ \pi t}{T}\right)}$, $(|g|, \Delta , \Omega_{0},\omega_{c})\times T=(60, 150, 60, 2416)$. As we can see from the figures, stronger is the leakage of the cavity less the state $P_{f,1}$ is populated.}
\label{fig:strong_weak_coupling}
\end{figure}

\begin{figure}[!ht]
         \includegraphics[width=0.45\textwidth]{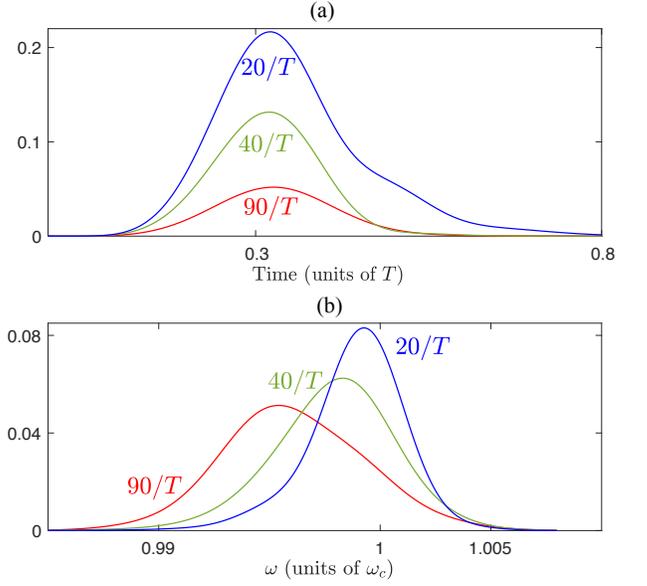}
         \caption{\small{Produced single photon shape in time (top figure) and frequency (bottom figure) domain for different values of the cavity decay rate $\Gamma_{c}$. The other parameters are the same as the ones in Fig.~\ref{fig:strong_weak_coupling}.}}
 \label{fig:different_regimes}        
\end{figure}

Here we start by analysing the dynamics for different coupling regimes {for a single mode cavity ($L=L_{0}$)}. We compare the strong coupling regime $g > \Gamma_{c}$ to an intermediate coupling regime $g\lesssim \Gamma_{c}$. {The parameters are chosen such that the approximate models described in Section~\ref{sec:model} remain valid.} Particular cases for intermediate and strong coupling regimes are presented in Fig.~\ref{fig:strong_weak_coupling}. As expected, in the strong coupling regime the single photon state inside the cavity is more populated than the one in an intermediate coupling regime. In Fig.~\ref{fig:different_regimes}, we study cavities of different decays and analyze the produced photon inside and outside the cavity, using the full inside-outside representation [Eq.\eqref{eq:dynamics_in_out}]. In Fig.~\ref{fig:different_regimes}~(a) the produced photon inside the cavity is presented ({$P_{f,1}$}). As one can expect, with the decrease of $\Gamma_{c}$ the probability of the photon state inside the cavity increases. We also notice that for the given symmetric $\Omega(t)$ the shape of the photon takes asymmetric form with the decrease of $\Gamma_{c}$. Fig.~\ref{fig:different_regimes}~(b) shows the shape of the leaked photon in frequency domain ({$P_{p,\rm{out}}$}). As we can see from the figure better is the cavity more the leaked photon is centred around the cavity resonance frequency. As expected, the bandwidth of the photon gets narrower with the decrease of $\Gamma_{c}$.

The direct control of production of the shape of a single leaking photon can be achieved for a large detuning $\Delta\gg \Omega, g$ (allowing the adiabatic elimination of the excited state $|e,\emptyset \rangle$ \cite{shore2011manipulating}) and an effective weak coupling regime: $\Gamma_c\gg G,g^2/\Delta$ with $G=-g\Omega/\Delta$ the (assumed positive) effective Raman coupling (allowing the adiabatic elimination of the state $|f,1\rangle$ (Fig.~\ref{fig:strong_weak_coupling})). 

The adiabatic eliminations lead to:
\begin{subequations}
\begin{eqnarray}
c_{g,0}(t)&=&e^{i\zeta(t)}e^{-\frac{\theta(t)}{2}},\\
\zeta(t)&=&\int_{t_i}^t dt'\frac{\Omega^2(t')}{\Delta},\\
\theta(t)&=&\int_{t_i}^t dt'\frac{4G^2(t')}{\Gamma_c}.\label{theta}
\end{eqnarray}
\end{subequations}
We denote the initial time $t_i=0$.
From $c_{g,0}(t)$, i.e. for given $g$, $\Delta$, and $\Omega(t)$, one can infer $c_{f,1}(t)=-i2(G(t)/\Gamma_c)c_{g,0,0}(t)$ and Eq. \eqref{flx1} then gives the shape of the photon flux:
\begin{equation}
\Phi(t)=\dot{\theta}(t)e^{-\theta(t)}.\label{flx2}
\end{equation}
The inverse calculation allows one to tailor a desired photon flux by deriving explicitly the corresponding $\Omega(t)$ (for given $g$ and $\Delta$). This is achieved by determining $\theta(t)$ from \eqref{flx2}:
\begin{equation}
\theta(t)=-\ln\left[1-\int_{0}^tdt'\Phi(t')\right].
\end{equation}
We get the simple expression for the Rabi frequency by deriving this latter equation and from \eqref{theta}:
\begin{equation}
\label{derivedRabi}
\Omega(t)=\frac{\Delta\sqrt{\Gamma_c}}{2g}\sqrt{\frac{\Phi(t)}{1-\int_0^t dt'\Phi(t')}}.
\end{equation}
We remark that this definition of the Rabi frequency can diverge at large time.
To prevent it, we introduce an efficiency parameter $\eta<1$ which will ensure that $\Omega(t\to+\infty)=0$ when $\Phi(t\to+\infty)=0$ \cite{Vasilev_2010,Nisbet_Jones_2011}.

\begin{figure*}[!ht]
\begin{center}
\includegraphics[scale=0.45]{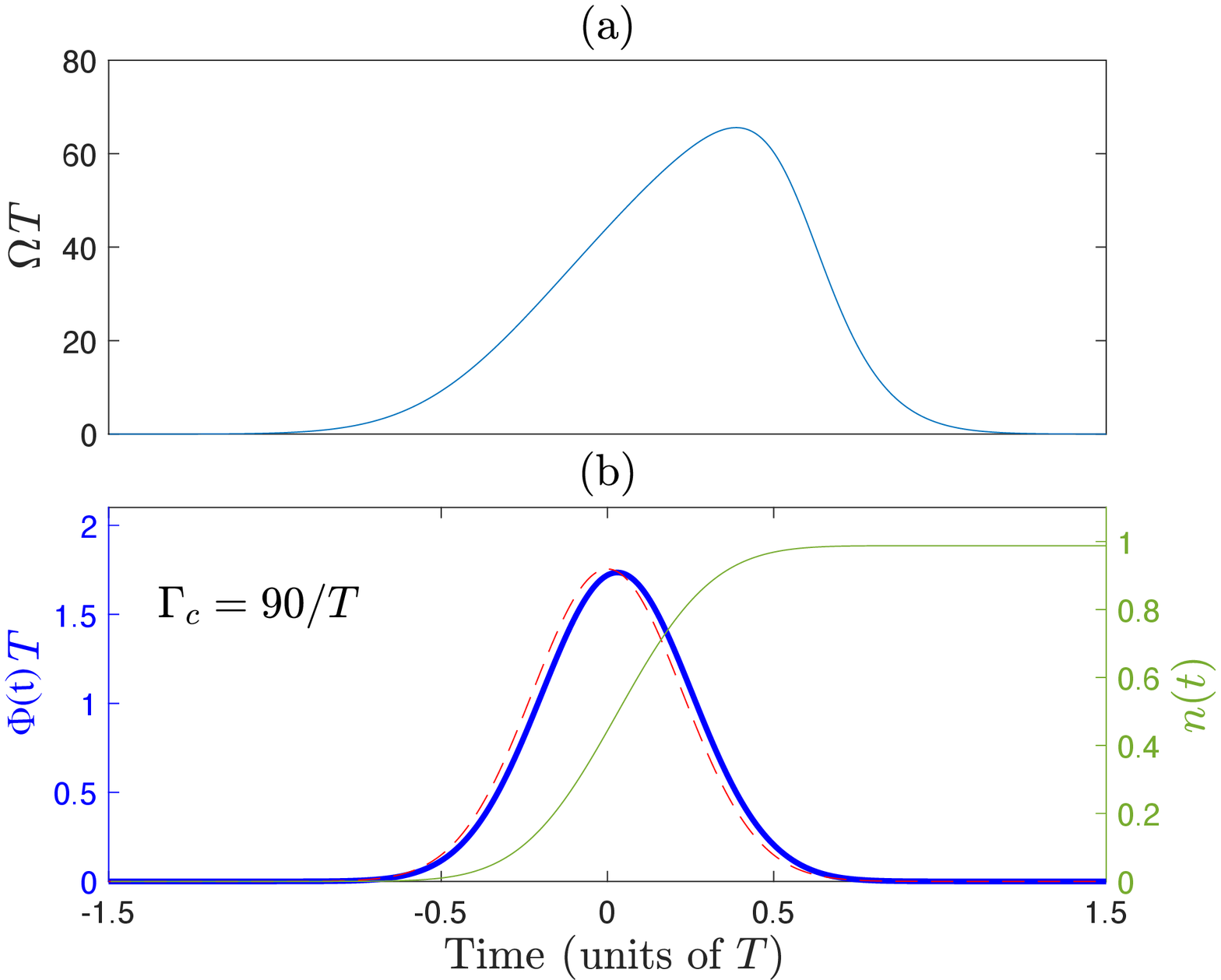}
\includegraphics[scale=0.45]{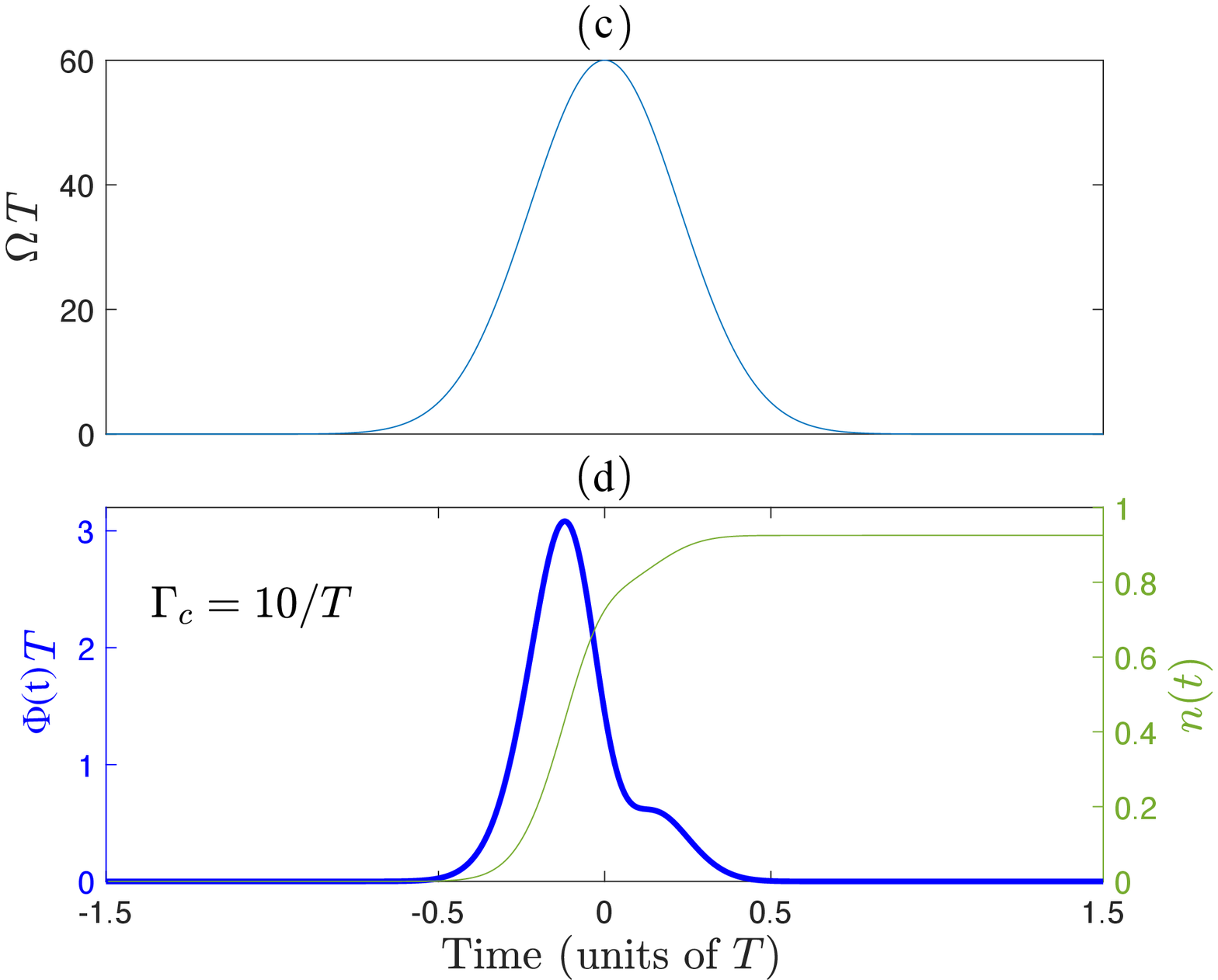}
\caption{\small \textcolor{black}{Left panel:} Rabi frequency $\Omega(t)T$ \eqref{derivedRabi} with $(|g|,\Gamma_c,\Delta)\times T=(60,90,300)$, $\eta=0.99$, determined from the desired Gaussian shape flux $\Phi(t)$ \eqref{phi_1} [desired (dashed line) and numerical from the original model \eqref{schroda_} (thick line)] of the single photon through the semi-transparent mirror (in units of $T$); number of outgoing photons $n=\int_{-\infty}^tdt'\Phi(t')=\Gamma_c\int_{-\infty}^tdt'|c_{f,1}(t')|^2$ during the process (thin line). \textcolor{black}{Right panel:} Same as above but for $\Gamma_c =10/T$ and a chosen Gaussian Rabi frequency $\Omega(t)=60 \exp[-\left(\pi t/T\right)^{2}]/T$.}
\label{photon-unique}
\end{center}
\end{figure*}

Numerical results for a chosen Gaussian probability for the single photon shape
\begin{equation}
\label{phi_1}
\Phi(t)=\frac{\eta \sqrt{\pi}}{T }\ e^{-\left(\frac{\pi t}{T}\right)^2},\quad \int_{-\infty}^{+\infty}\Phi(t) dt=\eta,
\end{equation}
are shown in Fig. \ref{photon-unique}\textcolor{black}{(a) and (b)}. Using $\Gamma_c=90/T$, we obtain $\max_tG(t)\approx 13/T\ll\Gamma_c$.
We have also checked numerically the resulting flux by determining it from the numerical solution of the Schr\"odinger equation \eqref{schroda_} (without considering the adiabatic elimination) with the Rabi frequency \eqref{derivedRabi}. The derived photon flux closely follows the desired shape as expected.

Other more complex forms can be investigated through \eqref{derivedRabi} {\cite{Reddy18,Krastanov2022,CiracDuanZoller}} such as the ones obtained by the resonant process with flying atoms in \cite{Vasilev_2010,Nisbet_Jones_2011}.

Fig.~\ref{photon-unique}\textcolor{black}{(c) and (d)} show a different situation with a cavity of better effective quality: $\Gamma_c=10/T$ and $\max_tG(t)= 12/T\approx\Gamma_c$, where the second adiabatic elimination cannot be made. In this case, the leakage of the photon occurs earlier and faster due to the earlier peak of the coupling. The smaller decay of the cavity leads to a deformation of the tail of the photonic shape.

\section{\label{sec:conclusion} Conclusion}

We have derived and analyzed models for a system of a $\Lambda$-atom trapped in a cavity, featuring a semi-transparent mirror, and driven by laser pulses {allowing the production of single photon leaking out from the cavity. We introduced true mode, inside-outside and pseudo-mode representations for describing the system from the first principles. From the exact modes of the system (the true mode representation), we explicitly introduce the cavity-reservoir coupling which allows one to describe the dynamics without any a priori approximations. We have demonstrated that under suitable approximations that we formulate, these different representations give accurate results that are similar to each other, yet generally differ. We particularly analyse a high-$Q$ cavity scenario and show that this requirement alone, in general, is not enough for these approximate models to work. This is especially significant for the models where we consider cavities with higher losses and mode overlaps, namely cavities with low refractive indices, such as plasmonic cavities. In nanophotonics, it is common to transpose these approximate models derived for optical cavities to plasmonic cavities. However, as shown here, these approximate models already yield different predictions for optical cavities with relatively high transmission.}

{In the literature, it is common to phenomenologically introduce the pseudo-mode representation. However, this kind of phenomenological approach does not provide the full description of the produced photon, namely the outgoing photon shape in frequency domain. In contrast, here, we recover the phenomenological model derived from the first principles; moreover, it is complemented with the complete description of the system, including the full characteristics of the photon in time as well as in  frequency domain.
This derivation justifies explicitly the Markov approximation producing an input-output relation from first principles via the non-trivial cavity-reservoir coupling \eqref{eq:kappa_coupling}. This allows a precise definition of the propagating outside photon state~\eqref{eq:out_photon}.}

Finally, concepts, such as Poynting vector, photon flux, input-output operators, photon state, that characterize the propagation of the resulting leaking photons, have been {defined and} connected: We have formulated an input-output relation taking into account the propagating effects, which allows a direct interpretation of the $b_{\text{out}}$ operator through the Poynting vector and the photon flux.
The generated flux is then determined from the quantum average of the dynamics of the photon number in cavity, which results from a standard master equation that we have derived using the operators at $x = 0$. Different coupling regimes have been discussed. In particular, we have studied an effective weak coupling regime with a large detuning and a strong cavity leakage, such that the adiabatic elimination of the cavity state is performed. In this case, one can directly link the envelope of the driving field to the  pulse shape of the outgoing single photon which can be tailored at will. 

In order to demonstrate the concepts in a straightforward way, we have considered a simple model for the mirror, as single layer with a fictitious large index. In practice, large index is produced via a multilayer mirror. Such more realistic model will be considered in future work. We will also take into account in a similar manner the reverse process of photon absorption and the full process of generation/absorption.

\begin{acknowledgments}

We acknowledge Hans Jauslin for helpful discussions. We acknowledge support from the European Union's Horizon 2020 research and innovation program under the Marie Sklodowska-Curie grant agreement No. 765075 (LIMQUET). This research was also supported by the Ministry of Culture and Innovation and the National Research, Development and Innovation Office within the Quantum Information National Laboratory of Hungary (Grant No. 2022-2.1.1-NL-2022-00004).

\end{acknowledgments}

\appendix
\color{black}{
\section{\label{app:delta_ev} Input output relation with the use of a Dirac delta  distribution}
In this Appendix, we show the mathematical inconsistency of the the standard development of the input output relation~\eqref{inout} from the Heisenberg-Langevin equations~\eqref{Heis-Lang2full} within the Markov approximation applied without model for the coupling (see, e.g., ~\cite{gardiner00}). By assuming a ``flat'' continuum via the approximation 
\begin{equation}
\kappa_c(\omega)\approx\sqrt{\frac{\Gamma_c}{2\pi}}
\end{equation}
and pushing the $\omega$ integration from $-\infty$,
we approximate the double integral in~\eqref{bdef_b} (considering for simplicity the case $x=0$) as
\begin{subequations}
\label{inputoutputdelta}
\begin{align}
\int_{t_0}^t dt'\int_0^{+\infty} d\omega\frac{\vert\kappa_c(\omega)\vert^2}{\sqrt{\Gamma_c}}c(t') e^{-i\omega(t-t')} \nonumber\\
\approx  \sqrt{\Gamma_c} \int_{t_0}^t dt' c(t')  \int_{-\infty}^{+\infty} \frac{d\omega}{2\pi} e^{-i\omega(t-t')} \\
= \sqrt{\Gamma_c} \int_{t_0}^t dt' c(t') \delta(t-t') \\
= \frac{\sqrt{\Gamma_c}}{2}  c(t).
\end{align}
\end{subequations}  
The last step, that can be reformulated in a simpler case as
\begin{align}
\label{delta12}
\int_{-\infty}^0 dt\, c(t) \delta(t) = \frac{1}{2}  c(0),
\end{align}
is mathematically undefined. Since the distributions are defined on the real line via the integration on a test function, it indeed necessitates the introduction of a multiplication with the Heaviside distribution:
\begin{align}
\int_{-\infty}^{+\infty} dt\, c(t) H(-t)\delta(t) = \int_{-\infty}^{0} dt\, c(t) \delta(t).
\end{align}
However such product of two non regular distributions is undefined, here more specifically the product of the Dirac delta distribution with the Heaviside distribution of discontinuity localized where the Dirac delta is infinite. This has been analyzed in Ref. \cite{Griffiths_delta}.

An explicit analysis can be conducted by defining a model for the Dirac delta distribution using a family of (non-even) functions represented in Fig. \ref{adelta}
\begin{align}
h_{\epsilon}^{a}(t)=\left\{
\begin{array}{ccc} 
0 & \text{for } & t<(a-1)\epsilon, \\
1/\epsilon & \text{for } & (a-1)\epsilon \le t \le a\epsilon, \\
0 & \text{for } & t>a\epsilon,
\end{array}
\right.
\end{align}
parametrized by a real number $0<a<1$,
in the limit $\epsilon\to0$. We can indeed check that they satisfy the Dirac delta distribution (applied on a test function $\varphi(t)$) : 
\begin{align}
\lim_{\epsilon\to0}\int_{-\infty}^{+\infty} dt\, \varphi(t) h_{\epsilon}^{a}(t)&=\lim_{\epsilon\to0}\frac{1}{\epsilon}\int_{(a-1)\epsilon}^{a\epsilon} dt\, \varphi(t),\nonumber\\
&=\lim_{\epsilon\to0} \int_{a-1}^{a} ds\, \varphi(\epsilon s),\nonumber\\
&= \varphi(0),
\end{align}
where we have applied the change of variable $s=t/\epsilon$. We note that the limit $\epsilon\to0$ guarantees that $\delta(t)=\lim_{\epsilon\to0}h_{\epsilon}^{a}(t)$ is an even distribution.

\begin{figure}[!h]
\begin{center}
  \begin{tikzpicture}[baseline]
\draw[thick,->] (-2,0) -- (2,0);
\draw[thick,->] (0,0) -- (0,2);
\draw(-1/3, 0) -- (-1/3,1) ;
\draw(-1/3, 1) -- (2/3,1) ;
\draw(2/3, 1) -- (2/3,0) ;
  \node at (0.5,1.8) {$h^a_\varepsilon(t)$};
    \node at (2,-0.5) {$t$};
        \node at (-3/3,-0.5) {$(a-1)\varepsilon$};
        \draw[->](-1/3, -0.3) -- (-1/3,-0.1) ;
         \node at (2.5/3,-0.5) {$a\varepsilon$};
        \draw[->](2/3, -0.3) -- (2/3,-0.1) ;
         \node at (3.2/3,1) {$1/\varepsilon$};
\end{tikzpicture}  
\caption{Representation of a family of non-even functions parametrized by a real number $0<a<1$ tending to the Dirac delta distribution in the limit $\epsilon\to0$.}
\label{adelta}
\end{center}
\end{figure}
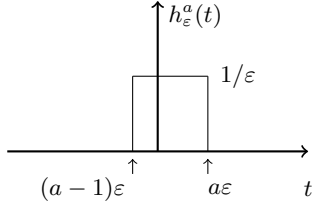

Applying this model on \eqref{delta12}, we obtain
\begin{align}
\lim_{\epsilon\to0}\int_{-\infty}^0 dt\, c(t) h_{\epsilon}^{a}(t) &=  \lim_{\epsilon\to0} \frac{1}{\epsilon}\int_{(a-1)\epsilon}^0 dt\, c(t) ,\nonumber\\
& =  \lim_{\epsilon\to0} \int_{a-1}^0 ds\, c(\epsilon s) ,\nonumber\\
&= (1-a) c(0).
\end{align}
This shows that the result depends on the details of the model of the Dirac delta distribution \cite{Saichev}. We recover the result of \eqref{delta12} only for a particular even-function model (i.e. $a=1/2$).   

As a consequence, the derivation \eqref{inputoutputdelta} is not valid in general. It necessitates a specific model for the coupling $\kappa_c(\omega)$, as it is considered from first principles in this paper.
}

{
\section{\label{app:Lorentzian_separation} Lorentzian structure of the cavity response function}

As it is shown in~\cite{dutra2005cavity, vogel2006quantum,multilayer}, for a one-dimensional, single layer dielectric cavity, having a perfect mirror placed at $x = -L$ and a semi-transparent mirror at $x=0$, the response function and the quantized electric filed inside the cavity write as follows:
\begin{eqnarray}
\label{eq:app:actual_resp}
T(\omega) &=& \frac{t(\omega)}{1+r(\omega)e^{2i\frac{\omega}{c}\left( L+\frac{\delta}{2}\right)}},\\ \nonumber
E_{\rm{in}}(x) &=& \int^{\infty}_{0} d\omega \sqrt{\frac{\hbar \omega}{\pi c \mathcal{A}\epsilon_{0}}}\sin{\left[\frac{\omega}{c}(x+L)\right]}e^{i\frac{\omega}{c}L}T(\omega)a_{\omega}\\ \label{eq:field_in}
&&+\,H.C.
\end{eqnarray}
where $t(\omega)$ and $r(\omega)$ are single layer spectral transmission and reflection functions:
\begin{eqnarray}
&&|t(\omega)|^{2}+|r(\omega)|^{2}=1,\\
&&t(\omega)r^{\ast}(\omega)+t^{\ast}(\omega)r(\omega)=0,
\end{eqnarray}
with
\begin{eqnarray}
t(\omega)&=&\frac{\left(1-r^{2}\right)e^{i(n-1)\frac{\omega}{c}\delta}}{1-e^{2in\frac{\omega}{c}\delta}r^{2}},\\
r(\omega)&=&e^{-i\frac{\omega}{c}\delta}\frac{r\left(e^{2in\frac{\omega}{c}\delta}-1\right)}{1-e^{2in\frac{\omega}{c}\delta}r^{2}}=|r(\omega)|e^{i\phi_{r}(\omega)},
\end{eqnarray}
 and $r=\frac{n-1}{n+1}$ is the reflectivity of the mirror with thickness $\delta = \frac{\lambda_{c}}{4n}$, $\lambda_{c}$ being the cavity resonance wavelength. It can be shown (\cite{dutra2005cavity, vogel2006quantum,multilayer}) that this response function can be written as a sum of Lorentzian-like functions, for a cavity with a low transmission rate:
 \begin{eqnarray}
 \label{eq:app:resp}
 |T(\omega)|^{2} \approx \sum_{m}\frac{c}{2L}\frac{\Gamma_{m}}{(\omega-\omega_{m})^{2}+\left(\frac{\Gamma_{m}}{2}\right)^{2}},
 \end{eqnarray}
 where 
 \begin{eqnarray}
 \Gamma_{m} &=& -\frac{c}{L}\ln{|r(\omega_{m})|},\\
 \label{eq:app:omega_m}
 \omega_{m} &=& m\frac{\pi c}{L}+\frac{c}{2L}\left( \pi - \phi_{r}(\omega_{m})\right).
 \end{eqnarray}
 For a high-$Q$ cavity with a low transmission rate ($|t(\omega)|\approx |\underline{t}|\ll1$, $\phi_{r}(\omega_{m})\approx \pi$), each Lorentzian of the sum~\eqref{eq:app:resp} is well separated from the others ($\Gamma_{m}\ll \pi c/L$), and we can deduce the following:
 \begin{eqnarray}
 \label{eq:lor_resp}
 T(\omega) = \sum_{m}\sqrt{\frac{c}{2L}}\frac{\sqrt{\Gamma_{m}}}{\omega-\omega_{m}+i\frac{\Gamma_{m}}{2}}.
 \end{eqnarray}
With~\eqref{eq:field_in} one can write the atom-environment interaction Hamiltonian from the dipole approximation~\cite{multilayer}: $H_{\rm{int}}=-d E_{\rm{in}}$, which,
 using the expression~\eqref{eq:lor_resp}, leads to the single mode coupling function of the form~\eqref{eq:coupling_full_Lor}.
 }
 \color{black}
 
\section{\label{app:integration} Discretization of the continuous integrals}

Eqs.~\eqref{eq:dynamics_full} and~\eqref{eq:dynamics_in_out} are integrated numerically via discretizing the continuum. {In order to do discretization properly, we analyse the continuous parts of the states~\eqref{eq:true_mode_state} and~\eqref{eq:state_in_out}, where the photon states $\lvert 1_{\omega}\rangle$ and $\lvert 1_{\omega,\rm{out}}\rangle$ and the coefficients $\tilde{c}_{f,1}(\omega,t)$ and $c_{f,0,1}(\omega,t)$ all have the unit of $1/\sqrt{\omega}$, making the corresponding integrals dimensionless. Thus we can do the following discretization:
\begin{eqnarray*}
&&\int^{\infty}_{0} \hspace{-1em}d\omega \,\tilde{c}_{f,1}(\omega,t)\lvert \mathbf{1}_{\omega}\rangle = \sum_{i=1}^{m} \sqrt{d\omega}\tilde{c}_{f,1}(\omega_{i},t) \sqrt{d\omega}\lvert \mathbf{1}_{\omega_{i}}\rangle,\\
&&\int^{\infty}_{0} \hspace{-1em}d\omega \, {c}_{f,0,1}(\omega,t)\lvert {1}_{\omega,\rm{out}}\rangle = \sum_{i=1}^{m} \sqrt{d\omega}{c}_{f,0,1}(\omega_{i},t) \sqrt{d\omega}\lvert {1}_{\omega_{i},\rm{out}}\rangle,
\end{eqnarray*}
where $d\omega$ is the step of the discretization. By denoting the dimensionless quantities of the sum as follows: $\mathbf{\tilde{c}}_{f,1}(\omega_{i},t) =\sqrt{{d}\omega}\,{\tilde{c}}_{f,1}(\omega_{i},t)$, $\lvert \tilde{\mathbf{1}}_{\omega_{i}}\rangle = \sqrt{d\omega}\lvert \mathbf{1}_{\omega_{i}}\rangle$ and $\mathbf{c}_{f,0,1}(\omega_{i},t)=\sqrt{d\omega}\,c_{f,0,1}(\omega_{i},t)$, $\lvert \bold{1}_{\omega_{i},\rm{out}}\rangle=\sqrt{d\omega}\lvert {1}_{\omega_{i},\rm{out}}\rangle$, we can calculate the probability amplitudes of finding the photon in states $\lvert \tilde{\mathbf{1}}_{\omega_{i}}\rangle$ and $\lvert \bold{1}_{\omega_{i},\rm{out}}\rangle$, respectively:
\begin{eqnarray*}
&&\tilde{P}(\omega_{i},t)=|\langle  \tilde{\mathbf{1}}_{\omega_{i}} \rvert \tilde{\psi} \rangle |^{2}=|\mathbf{\tilde{c}}_{f,1}(\omega_{i},t)|^{2},\\
&& P(\omega_{i},t) = |\langle  \bold{1}_{\omega_{i},\rm{out}} \rvert \psi \rangle|^{2} = |\mathbf{c}_{f,0,1}(\omega_{i},t)|^{2}.
\end{eqnarray*}
Taking this discretization into account, in Eq.~\eqref{eq:dynamics_full}}, the discretization of the function $\eta(\omega)$ becomes $\sqrt{d\omega}\,\eta(\omega_{1}), \sqrt{d\omega}\,\eta(\omega_{2}), \cdots \sqrt{d\omega}\,\eta(\omega_{m})$, and the equations become: 
\begin{subequations}
\begin{eqnarray*}
i\dot{\tilde{c}}_{g,0}(t)& =& \Omega \, \tilde{c}_{e,0}(t),\\
i\dot{\tilde{c}}_{e,0}(t)& =& \Delta \, \tilde{c}_{e,0}(t)+\Omega \, \tilde{c}_{g,0}(t)+ i\sum_{m} \; \tilde{\eta}_{\omega_{m}}\tilde{\mathbf{c}}_{f,1}(\omega_{m},t),\\
i\dot{\tilde{\mathbf{c}}}_{f,1}(\omega_{1},t)& = &\left( \Delta-\Delta_{c}+\omega_{1} - \omega_{c}\right)\mathbf{\tilde{c}}_{f,1}(\omega_{1},t)-i \tilde{\eta}^{\ast}_{\omega_{1}} \tilde{c}_{e,0}(t),\\
& \vdots &\\
i\dot{\tilde{\mathbf{c}}}_{f,1}(\omega_{m},t)& = &\left( \Delta-\Delta_{c}+ \omega_{m} - \omega_{c}\right)\mathbf{\tilde{c}}_{f,1}(\omega_{m},t)-i \tilde{\eta}^{\ast}_{\omega_{m}} \tilde{c}_{e,0}(t),
\end{eqnarray*}
\end{subequations}
with $\tilde{\eta}_{\omega_{m}}= \sqrt{d\omega}\,\eta(\omega_{m})$.

Similarly, Eq.~\eqref{eq:dynamics_in_out} becomes: 
\begin{subequations}
\begin{eqnarray*}
i\dot{c}_{g,0}(t) &=& \Omega \,c_{e,0}(t),\\
i\dot{c}_{e,0}(t) &=& \Delta \, c_{e,0}(t)+\Omega \, c_{g,0}(t)+g\, c_{f,1,0}(t),\\ 
i\dot{c}_{f,1,0}(t) &=&\left( \Delta-\Delta_{c}\right)c_{f,1,0}+ g\, c_{e,0}(t)\\
&{}&-\,i\sum_{m}  \tilde{\kappa}_{c}^{\ast}(\omega_{m})\,\mathbf{c}_{f,0,1}(\omega_{m},t),\\
i\dot{\mathbf{c}}_{f,0,1}(\omega_{1},t) &=& \left(  \Delta-\Delta_{c}+\omega_{1} - \omega_{c}\right)\mathbf{c}_{f,0,1}(\omega_{1},t)\\
&{}&+ \, i \tilde{\kappa}_{c}(\omega_{1}) \,c_{f,1,0}(t),\\
&\vdots&\\
i\dot{\mathbf{c}}_{f,0,1}(\omega_{m},t) &=& \left( \Delta-\Delta_{c}+ \omega_{m} - \omega_{c}\right)\mathbf{c}_{f,0,1}(\omega_{m},t)\\
&{}&+\, i \tilde{\kappa}_{c}(\omega_{m}) \,c_{f,1,0}(t),
\end{eqnarray*}
\end{subequations}
where $\tilde{\kappa}_{c}(\omega_{m})=\sqrt{d\omega}\, \kappa_{c}(\omega_{m})$.

By solving these systems of equations numerically using a sufficiently large number of (typically 100000) states discretizing the continuum, we obtain the solutions presented in Fig.~\ref{equivalence}.

\section{\label{app:poynting} Poynting vector derivation}

Following the definition of Poynting vector~\cite{dutra2005cavity,Blow_1990} we can write it in the true mode representation, using the modes derived for the outside of the cavity \cite{dutra2005cavity,vogel2006quantum, multilayer}:
\begin{eqnarray*}
S(x)&=& -\frac{1}{2\mu_{0}}(B_{\rm{out}}(x)E_{\rm{out}}(x)+E_{\rm{out}}(x)B_{\rm{out}}(x))
\end{eqnarray*} 
where $\mu_{0} = 1/(c^{2}\epsilon_{0})$ is the vacuum permeability and 
\begin{eqnarray*}
E_{\rm{out}}(x) &=& -\frac{i}{\sqrt{2\pi c\mathcal{A}}} \int_{0}^{\infty} \hspace{-1em} d\omega \sqrt{\frac{\hbar \omega}{2\epsilon_{0}}}\Big(\left(R_{\omega}e^{i\frac{\omega}{c}x}-e^{-i\frac{\omega}{c}x}\right)a_{\omega}\\
&-& H.C.\Big),\\
B_{\rm{out}}(x) &=& \frac{i}{c\sqrt{2\pi c\mathcal{A}}} \int_{0}^{\infty} \hspace{-1em} d\omega \sqrt{\frac{\hbar \omega}{2\epsilon_{0}}}\Big(\left(R_{\omega}e^{i\frac{\omega}{c}x}+e^{-i\frac{\omega}{c}x}\right)a_{\omega}\\
&-& H.C.\Big),
\end{eqnarray*}
with 
\begin{eqnarray*}
R_{\omega}=e^{2i\frac{\omega}{c}L}\frac{T(\omega)}{T^{\ast}(\omega)}\approx \sqrt{\frac{2\pi}{\Gamma_{c}}}\sqrt{\frac{\omega_{c}}{\omega}}\alpha^{\ast}(\omega)\Big(\omega-\omega_{c}-i\frac{\Gamma_{c}}{2}\Big).
\end{eqnarray*}
$\alpha(\omega)$ is the coefficient linking the true mode $a_{\omega}$ to the discrete cavity mode $c$: $a_{\omega} = \alpha(\omega) c+\int^{\infty}_{0} d\omega' \beta(\omega,\omega') b_{\omega'}$, and can be written as follows \cite{dutra2005cavity}:
\begin{eqnarray*}
\alpha(\omega)&=&\sqrt{\frac{\omega}{\omega_{c}}}\sqrt{\frac{L}{\pi c}}\text{sinc} \left((\omega-\omega_{c})\frac{L}{c}\right) e^{-i\frac{\omega}{c}L}T^{\ast}(\omega)
\end{eqnarray*}
Taking these into account we get the following Poynting vector, for the propagation in the positive $x$ direction:
\begin{eqnarray*}
S(x)&=& \frac{\hbar}{2\pi \mathcal{A}}\int^{\infty}_{0} \hspace{-1em}d\omega d\omega' \sqrt{\omega\omega'}{\rm Re}\Big\{R_{\omega}R^{\ast}_{\omega'}e^{i\frac{(\omega-\omega')}{c}x}a_{\omega'}^{\dagger}a_{\omega}\Big\}.
\end{eqnarray*}
Further, by writing $a_{\omega}$ in terms of the outside operator corresponding to the inside-outside representation: $a_{\omega} = \int d\omega' \beta(\omega,\omega')b_{\omega'}$, and recalling that $i\int d\omega d\omega' \omega  \alpha^{\ast}(\omega)\beta(\omega,\omega') b_{\omega'} = \int d\omega \kappa_{c}^{\ast}(\omega)b_{\omega}$ \cite{dutra2005cavity}, we obtain the following:
\begin{eqnarray*}
S(x)&=& \frac{\hbar }{ \mathcal{A}}\frac{\omega_{c}}{\Gamma_{c}}\int^{\infty}_{0} \hspace{-1em}d\omega d\omega'  \kappa_{c}^{\ast}(\omega) \kappa(\omega')e^{i\frac{(\omega-\omega')}{c}x} b^{\dagger}_{\omega'}b_{\omega}.
\end{eqnarray*}
Via defining the integrated reservoir operator as follows:
\begin{eqnarray}
b(x) = \frac{1}{\sqrt{\Gamma_{c}}}\int^{\infty}_{0} \hspace{-1em}d\omega \kappa^{\ast}(\omega) e^{i\frac{\omega}{c}x}b_{\omega},
\end{eqnarray}
the expression for Poynting vector becomes
\begin{eqnarray}
S(x)&=& \frac{\hbar }{ \mathcal{A}}\frac{\omega_{c}}{\Gamma_{c}}b^{\dagger}(x)b(x).
\end{eqnarray}
\color{black}
\section{\label{app:evaluation_of_integral} Evaluation of the integral~\eqref{bdef_b}}

In order to evaluate the integral in~\eqref{bdef_b} we use the following expression for $\kappa_{c}(\omega)$:
\begin{align*}
\kappa_{c}(\omega) = \sqrt{\frac{\Gamma_{c}}{2\pi}}e^{-i\frac{\omega}{c}L}\text{sinc}{\left ( (\omega-\omega_{c})\frac{L}{c}\right)}.
\end{align*}
We calculate the following integral:
\begin{align}
\nonumber
\int_{0}^{\infty} d\omega \; |{\kappa_{c}(\omega)}|^{2}e^{-i\omega \tau}&= \frac{{\Gamma_{c}}}{2\pi}\left(\frac{c}{L}\right)^{2}\\ \label{app:integral}
&\times \int_{0}^{\infty} d\omega \; \frac{\sin^{2}{\left ( (\omega-\omega_{c})\frac{L}{c}\right)}}{(\omega-\omega_{c})^{2}}e^{-i\omega \tau} ,
\end{align} 
where $\tau = t-t'-\frac{x}{c}$. This leads to the evaluation of the integral of the following form (\textcolor{black}{since $\omega_{c}=\frac{\pi c}{L}$, this is significantly larger than $\frac{c}{L}$, such that the integral can be evaluated as: $\int_{0}^{\infty}\rightarrow \int_{-\infty}^{\infty}$}):
\begin{align*}
\int^{\infty}_{-\infty} dx \; \frac{e^{\pm ix\tau}}{x^{\textcolor{black}{2}}}.
\end{align*} 
This integral can be evaluated in the complex plane, and its value is different for negative and positive parameters $\tau$. Taking this into account it can be shown that the integral in~\eqref{app:integral} becomes 
\begin{align}
\nonumber
&\frac{\Gamma_{c}}{2\pi}\left(\frac{c}{L}\right)^{2}\frac{1}{2}e^{-i\omega_{c}(t-t'-\frac{x}{c})} \\ \label{eq:app:integral_value}
&\times \begin{cases}
\begin{array}{c c}
0, & t'>t-\frac{x}{c}+\frac{2L}{c}\\
-\pi(t'-t+\frac{x}{c}-\frac{2L}{c}), & t-\frac{x}{c}<t'<t-\frac{x}{c}+\frac{2L}{c}\\
\pi(t'-t+\frac{x}{c}+\frac{2L}{c}), & t-\frac{x}{c}-\frac{2L}{c}<t'<t-\frac{x}{c}\\
0, & t'<t-\frac{x}{c}-\frac{2L}{c}
\end{array}\end{cases}
\end{align}
Having calculated the integral over the frequency, we can now evaluate the time integral in~\eqref{bdef_b}. Taking into account the results in~\eqref{eq:app:integral_value}, we can reduce the integration range to the following ones
\begin{align*}
&\int_{t_{0}}^{t}  = \int_{t-\frac{x}{c}-\frac{2L}{c}}^{t-\frac{x}{c}}+\int_{t-\frac{x}{c}}^{t-\frac{x}{c}+\frac{2L}{c}}, \qquad {\text{when}~ x>2L},\\
&{\int_{t_{0}}^{t}  = \int_{t-\frac{x}{c}-\frac{2L}{c}}^{t-\frac{x}{c}}+\int_{t-\frac{x}{c}}^{t}, \qquad \text{when}~ 0<x<2L}\\
\end{align*}
with $t>t_{0}+\frac{x}{c}+\frac{2L}{c}$.
Considering that we analyse the dynamics for the times much bigger than the round trip time of the produced photon, i.e. $t\gg \frac{2L}{c}$ (coarse-grained approximation),  {the integral for the case $x<2L$ can be evaluated the same way as the integral for $x>2L$, since $t +2L/c > t-x/c +2L/c \approx t$}. Hence, the integrals above can be evaluated as follows: 
\begin{align*}
\int^{t-\frac{x}{c}+\frac{2L}{c}}_{t-\frac{x}{c}}\hspace{-0.75em} dt' f(t') &=\frac{1}{2}\Big[f\Big(t-\frac{x}{c}+\frac{2L}{c}\Big)+f\left(t-\frac{x}{c}\right)\Big] \frac{2L}{c},\\
\int^{t-\frac{x}{c}}_{t-\frac{x}{c}-\frac{2L}{c}} \hspace{-0.75em} dt' f(t') & = \frac{1}{2}\Big[f\Big(t-\frac{x}{c}-\frac{2L}{c}\Big)+f\left(t-\frac{x}{c}\right)\Big] \frac{2L}{c}.
\end{align*}
Hence, for the full integral we obtain:
\begin{align*}
\int_{t_{0}}^{t} dt' \int_{0}^{\infty} d\omega \; |{\kappa_{c}(\omega)}|^{2}e^{-i\omega (t-t'-\frac{x}{c})}{c}(t')&= \Gamma_{c} {c}\left(t-\frac{x}{c}\right). 
\end{align*}
For the case $x = 0$, the integration over the frequency in~\eqref{app:integral} gives the following result:
\begin{align}
\nonumber
&\frac{\Gamma_{c}}{2\pi}\left(\frac{c}{L}\right)^{2}\frac{1}{2}e^{-i\omega_{c}(t-t')}\\ \label{eq:app:integral_value_z0}
&\times \begin{cases}
\begin{array}{c c}
0, & t'>\frac{2L}{c}\\
-\pi(t'-t+\frac{x}{c}-\frac{2L}{c}), & t<t'<t+\frac{2L}{c}\\
\pi(t'-t+\frac{x}{c}+\frac{2L}{c}), & t-\frac{2L}{c}<t'<t\\
0, & t'<t-\frac{2L}{c}
\end{array}\end{cases}.
\end{align}
 Since the upper limit of the time integration is $t$, the second line of \eqref{eq:app:integral_value_z0} does not contribute to the integration over the time and the overall integral becomes: 
 \begin{align*}
\int_{t_{0}}^{t} dt' \int_{0}^{\infty} d\omega \; |{\kappa_{c}(\omega)}|^{2}e^{-i\omega (t-t')}{c}(t')&= \frac{\Gamma_{c}}{2} {c}\left(t\right). 
\end{align*}

\section{\label{app:master_eq} Derivation of the master equation}

In the following, we derive the dynamics of $X_{S}(t)$ from the Heisenberg equation, using Eqs.~\eqref{Hamiltonian} and~\eqref{eq:HSyst}:
\begin{eqnarray}
\nonumber
\frac{d}{dt}X_{S}(t) & = & -\frac{i}{\hbar}[X_{S}(t),H^{(H)}(t)] \, = -\frac{i}{\hbar}[X_{S}(t),H_{S}^{(H)}(t)]\\
 & {} &+ \, \int_{0}^{\infty}d\omega\Big(\kappa_{c}(\omega)b_{\omega}^{\dagger}(t)\left[X_{S}(t),c(t)\right]\\
 \nonumber
 & {} &-\, \kappa_{c}^{\ast}(\omega)\left[X_{S}(t),c^{\dagger}(t)\right]b_{\omega}(t)\Big).
 \end{eqnarray} 
 From the definition \eqref{bdef}, we have $\int_{0}^{\infty}d\omega \, \kappa_{c}^{\ast}(\omega)b_{\omega}(t) =\sqrt{\Gamma_{c}} {b}(z=0,t)$, for which we can use the relation \eqref{bath0}; hence 
 \begin{eqnarray*}
&& \frac{d}{dt}X_{S}(t)  =  -\frac{i}{\hbar}[X_{S}(t),H_{S}^{(H)}(t)]\\
 &&+ \, \Big(\sqrt{\Gamma_{c}}b_{in}^{\dagger}(t)+\frac{\Gamma_{c}}{2}c^{\dagger}(t)\Big)\left[X_{S}(t),c(t)\right]\\
 &&- \, \left[X_{S}(t),c^{\dagger}(t)\right]\Big(\sqrt{\Gamma_{c}}b_{in}(t)+\frac{\Gamma_{c}}{2}c(t)\Big)\\
 &&=- \frac{i}{\hbar}[X_{S}(t),H_{S}^{(H)}(t)]\\
 &&+\, \sqrt{\Gamma_{c}} b_{in}^{\dagger}(t)\left[X_{S}(t),c(t)\right] -\left[X_{S}(t),c^{\dagger}(t)\right]\sqrt{\Gamma_{c}}b_{in}(t)\\
 && + \, \Gamma_{c}\Big(c^{\dagger}(t)X_{S}(t)c(t)-\frac{1}{2}\left\{ c^{\dagger}(t)c(t),X_{S}(t)\right\} \Big). 
 \end{eqnarray*}
 We further define the time-dependent dissipator $\mathcal{D}^{\dagger}_{\text{in},t}(X_{S}(t))= \sqrt{\Gamma_{c}}b_{in}^{\dagger}(t)\left[X_{S}(t),c(t)\right] -\left[X_{S}(t),c^{\dagger}(t)\right]\sqrt{\Gamma_{c}}b_{in}(t)$, leading to 
\begin{eqnarray}
 \label{eq:dynamic_xs}
 \frac{d}{dt}X_{S}(t) & = & -\frac{i}{\hbar}[X_{S}(t),H_{S}^{(H)}(t)]+\mathcal{D}^{\dagger}_{\text{in},t}(X_{S}(t))\\
  \nonumber
 &{}&+\, \Gamma_{c}\Big(c^{\dagger}(t)X_{S}(t)c(t)-\frac{1}{2}\left\{ c^{\dagger}(t)c(t),X_{S}(t)\right\} \Big)
 \end{eqnarray}
 The expectation value of $X_{S}$ can be calculated as follows:
   \begin{eqnarray*}
  \langle X_{S}(t)\rangle &=& \text{Tr} \left \{ X_{S}(t)\rho (t_{0})\right\} \\
  &=&\text{Tr} \left \{ X_{S}U(t,t_{0})\rho (t_{0})U^{\dagger}(t,t_{0})\right\}\\
  &=&\text{Tr}_{S} \left \{\text{Tr}_{R} \left \{X_{S}U(t,t_{0})\rho (t_{0})U^{\dagger}(t,t_{0}) \right\} \right \}\\
  &=&\text{Tr}_{S} \left \{X_{S}\rho_{S} (t) \right\},
  \end{eqnarray*}
  where we have used the cyclic property of the trace, and defined $\rho_{S}(t)=\text{Tr}_{R}\left \{ U(t,t_{0})\rho (t_{0})U^{\dagger}(t,t_{0}) \right\}$. Similarly, using the property $\text{Tr}\left \{ A+B\right \}=\text{Tr}\left \{ A\right \} + \text{Tr}\left \{ B\right \}$, $\forall \, A, \,B$, we can calculate the averages on the right hand side of Eq.~\eqref{eq:dynamic_xs}:
\begin{align*}
\big \langle  &[ X_{S}(t),H_{S}^{(H)}(t)] \big \rangle =  \text{Tr}\left\{ [ X_{S}(t),H_{S}^{(H)}(t)]\rho(t_{0})\right\} \\
& =  \text{Tr}\left\{ \left[ X_{S},H_{S}(t)\right]U(t,t_{0})\rho(t_{0})U^{\dagger}(t,t_{0})\right\}  \\
 & =   \text{Tr}_{S}\left\{ \left[ X_{S},H_{S}(t)\right]\rho_{S}(t)\right\} =  \text{Tr}_{S}\left\{ X_{S} \left[H_{S}(t),\rho_{S}(t)\right]\right\},
 \end{align*}
 \begin{align*}
  \big \langle  c^{\dagger}(t)X_{S}(t)c(t) \big \rangle &= \text{Tr}_{S}\left \{ c^{\dagger}X_{S}c \rho_{S}(t)\right \}\\
  &= \text{Tr}_{S}\left \{ X_{S}c \rho_{S}(t)c^{\dagger}\right \},\\
  \big \langle \{ c^{\dagger}(t)c(t),X_{S}(t)\} \big \rangle \, &= \text{Tr}_{S}\left \{\left\{ c^{\dagger}c,X_{S}\right\}\rho_{S}(t)\right \} \\
&=  \text{Tr}_{S}\left \{ X_{S} \left\{ \rho_{S}(t),c^{\dagger}c\right\}\right \}.
 \end{align*}
Assuming that the reservoir is initially a vacuum state $\rho_{R}(t_{0})=\lvert \emptyset \rangle \langle \emptyset \lvert$, for the dissipator part $\mathcal{D}^{\dagger}_{\text{in},t}$ we get
\begin{eqnarray*}
\text{Tr} \{ b_{in}^{\dagger}(t)&&\left[X_{S}(t),c(t)\right] \rho (t_{0}) \} \\
&&=\text{Tr} \{ \left[X_{S}(t),c(t)\right] \rho (t_{0}) b_{in}^{\dagger}(t)\}\\
&&=\text{Tr} \{ \left[X_{S}(t),c(t)\right] \rho_{S} (t_{0})\otimes \rho_{R}(t_{0}) b_{in}^{\dagger}(t)\}\\
&&=\text{Tr} \{ \left[X_{S}(t),c(t)\right] \rho_{S} (t_{0})\otimes \lvert \emptyset \rangle \langle \emptyset \lvert b_{in}^{\dagger}(t)\} =0.
\end{eqnarray*}
Similarly
\begin{eqnarray*}
&&\text{Tr}\left \{ \left[X_{S}(t),c^{\dagger}(t)\right]b_{in}(t) \rho (t_{0}) \right\} =0.
\end{eqnarray*}
Finally, Eq.~\eqref{eq:dynamic_xs} becomes 
\begin{eqnarray*}
&&\text{Tr}_{S}\Big\{ X_{S}\frac{d\rho_{S}(t)}{dt}\Big \}   =  \text{Tr}_{S}\left\{ X_{S}\left[H_{S}(t),\rho_{S}(t)\right]\right\}\\
&&{}\, + \,  \Gamma_{c}\Big(\text{Tr}_{S}\left\{ X_{S}c\rho_{S}(t)c^{\dagger}\right\} -\frac{1}{2}\text{Tr}_{S}\left\{ X_{S}\left\{ \rho_{S}(t),c^{\dagger}c\right\} \right\} \Big).
\end{eqnarray*}
Further, using the property $\forall A,   \, \text{Tr} \left \{ AB \right \} = \text{Tr}\left \{ AC \right \} \Leftrightarrow B=C$, we obtain the master equation for $\rho_{S}(t)$:
\begin{eqnarray*}
\frac{d}{dt} \rho_{S}(t)& = & \left[H_{S}(t),\rho_{S}(t)\right]+\Gamma_{c}\Big(c\rho_{S}(t)c^{\dagger}-\frac{1}{2}\left\{ \rho_{S}(t),c^{\dagger}c\right\} \Big).
\end{eqnarray*}

\bibliography{cav_photon}

\end{document}